\documentclass[12pt,letterpaper]{article}
\pdfoutput=1
\usepackage[colorlinks=true,
linkcolor=Blue, 
citecolor=Blue,
filecolor=Blue,
urlcolor=Blue,
linktoc=page, 
pdfstartview=FitV,
bookmarksopen=true]{hyperref}
\usepackage[left=2cm,top=1cm,right=3cm,nohead]{geometry}
\usepackage[dvipsnames]{xcolor}
\usepackage{colortbl,cancel}
\usepackage[utf8]{inputenc} 
\usepackage{color}
\usepackage{mcite}
\usepackage{epsfig,latexsym,amsfonts,mathtools,amsthm,amssymb,amsbsy,multirow,slashed,
	mathrsfs,wasysym,textcomp,subfigure,wrapfig,comment,bbold,array,longtable,multirow,setspace,amsmath,pifont,nicefrac,footmisc}
\usepackage{tabularx}
\usepackage[]{mdframed}
\usepackage{graphicx}
\usepackage{setspace}
\usepackage{subfigure}
\usepackage{enumitem}
\usepackage{multirow}
\usepackage[bf]{caption}
\usepackage{braket}
\usepackage{amssymb}
\usepackage{comment}
\usepackage{float}
\usepackage{tikz}
\usetikzlibrary{decorations.pathreplacing}

\colorlet{color1}{gray!25}
\usetikzlibrary{positioning}
\usetikzlibrary{intersections} 
\newlength{\PicScale}
\usepackage[UKenglish]{babel}
\usepackage[toc,page]{appendix}
\usepackage{hhline}
\usepackage{geometry}
\usepackage{cite}
\usepackage{datetime}
\usepackage{bbm} 
\usepackage{bm} 
\hypersetup{
	pdftitle={},
	pdfauthor={},
	pdfsubject={}
}

\newcolumntype{M}[1]{>{\centering\arraybackslash}m{#1}}
\newcolumntype{N}{@{}m{0pt}@{}}

\numberwithin{equation}{section}

\newcommand{\Z}{\mathbb{Z}}

\makeatletter
\def\@cline#1-#2\@nil{
	\omit
	\@multicnt#1
	\advance\@multispan\m@ne
	\ifnum\@multicnt=\@ne\@firstofone{&\omit}\fi
	\@multicnt#2
	\advance\@multicnt-#1
	\advance\@multispan\@ne
	\leaders\hrule\@height\arrayrulewidth\hfill
	\cr
	\noalign{\nobreak\vskip-\arrayrulewidth}}
\makeatother

\begin{document}
	\pagestyle{empty}
	\begin{center}        
		{\bf\LARGE String islands, discrete theta angles and \\the 6D $\mathcal{N} = (1,1)$ string landscape\\ [3mm]}
		
        
		\large{Zihni Kaan Baykara$^1$, H\'ector Parra De Freitas$^1$ and Houri-Christina Tarazi$^{2,3}$
			\\[2mm]}

		{\small ${}^1$ Jefferson Physical Laboratory, Harvard University\\ [-1mm]}
		{\small\textit{Cambridge, MA 02138, USA}\\[0.2cm]}
        {\small ${}^2$ Enrico Fermi Institute \& Kadanoff Center for Theoretical Physics, University of Chicago\\ [-1mm]}
		{\small\textit{Chicago, IL 60637, USA}\\[0.2cm]}

 {\small ${}^3$ Kavli Institute for Cosmological Physics, University of Chicago\\ [-1mm]}
 {\small\textit{Chicago, IL 60637, USA}\\[0.2cm]}

		{\small \verb"zbaykara@g.harvard.edu, hparradefreitas@fas.harvard.edu, htarazi@uchicago.edu"\\[-3mm]}
		\vspace{0.3in}

		\small{\bf Abstract} \\[3mm]\end{center}
        The complete classification of the landscape of 6D $\mathcal{N} = (1,1)$ string vacua remains an open problem. In this work we prove a classification theorem for 6D $\mathcal{N} = (1,1)$ asymmetric orbifolds utilizing a correspondence with orbifolds of chiral 2D SCFTs with $c= 24$ (or $c= 12$). Interestingly, this class of theories can give rise to 6D vacua in which the only massless degrees of freedom reside in the gravity multiplet, with no moduli other than the dilaton—thus corresponding to truly isolated vacua, called \textit{string islands}. It is expected that there exist five new type II islands with as-yet-unknown constructions. In this work we construct them all using asymmetric $\mathbb{Z}_n$-orbifolds of Type II on $T^4$ with $n = 5,8,10,12$. We show that the cases $n = 5,8$ admit non-trivial discrete theta angles which have important consequences for both the string and particle charge lattices. In fact they provide examples of BPS-incompleteness and the strongest failure of the lattice weak gravity conjecture. Our work is expected to finalize our understanding of all perturbative 6D $\mathcal{N} = (1,1)$ theories.
\newpage



\setcounter{page}{1}
\pagestyle{plain}
\renewcommand{\thefootnote}{\arabic{footnote}}
\setcounter{footnote}{0}

\tableofcontents	

\newpage
\section{Introduction}\label{s:intro}

Despite decades of research in string theory, its vast landscape of vacua remains largely unexplored. In particular, large classes of such compactifications have been constructed using diverse techniques such as string orbifolds, orientifolds and  Calabi-Yau compactifications, but it remains unclear how close we are to achieving a full description in a given regime (e.g. $D$-dimensional Minkowski spacetime with $N$ supersymmetries). Luckily, this landscape is expected to be finite \cite{Yau:1991,acharya2006finitelandscape,Hamada_2022, vafa2005stringlandscapeswampland} making an exhaustive search manageable, especially when a high amount of supersymmetry is present. In this work we focus on addressing this issue in the presence of half-maximal supersymmetry $N=16$, with low energy degrees of freedom  residing in the gravity multiplet or  vector multiplets. The rank of such theories is bounded by $r_G\leq 26-D$, making their classification particularly manageable.

A renewed interest in the complete charting of the string landscape was reinforced by the recent progress of the Swampland program to demonstrate finiteness \cite{Kim:2019ths} and string universality in $D \geq 7$ with 16 supercharges\cite{Cveti__2020, Bedroya:2021fbu,Hamada:2021bbz,bedroya2023nonbpspathstringlamppost}. In particular, the ranks fall into the following possibilities: 
\begin{equation*}
D = 9: ~~~~~ r=17, 9, 1 ~~~~~\,; ~~~~~ D = 8: ~~~~~ r=18, 10, 2 ~~~~~\,; ~~~~~ D = 7: ~~~~~ r= 19, 11, 7, 5, 3, 1\,.
\end{equation*}

For example, it was argued from the bottom-up in \cite{Bedroya:2021fbu} that the 7D sector of the quantum gravity landscape should be completely described by M-theory compactified on a K3 surface up to 3-form fluxes freezing the blow-up moduli of suitable ADE singularities. This matches the top-down ideas of \cite{deBoer:2001wca} where such backgrounds were introduced, and conjectured to exist under the assumption that M-theory should describe every 7D string vacuum in some region of its moduli space.

Moreover, a systematic scan of frozen singularity backgrounds of K3 \cite{ParraDeFreitas:2022wnz} determined that three different ADE singularity configurations can each be realized in two inequivalent ways, and this in turn predicted the existence of three new 7D stringy vacua which were found to be realized by turning on discrete theta angles \cite{Montero:2022vva} in three different type II compactifications. Moreover, one of these vacua can be nontrivially decompactified to 9D, and so discrete theta angles extend the string landscape in $D = 7,8,9$.

However, the 6d $\mathcal{N}=(1,1)$ stringy landscape remains incomplete, hence motivating this work\footnote{Note that the chiral 6d theory is unique due to anomalies and can be realized as type IIB on $K3$.}. A way to realize a large class of these theories is using \textit{asymmetric orbifolds} with chiral actions\footnote{By chiral actions we mean orbifold actions acting exclusively on the worldsheet right-movers or left-movers, excluding quasicrystalline actions \cite{Harvey:1987da} which were classified in \cite{Baykara:2024vss}.} \cite{Narain:1986qm} of either heterotic or type II strings, with the latter sometimes admitting nonperturbative \textit{discrete theta angles} \cite{Montero:2022vva}. For $\mathbb{Z}_n$-orbifolds, it was conjectured in \cite{Fraiman:2022aik} that, at the worldsheet level (i.e. up to discrete theta angle), the heterotic strings correspond to families of currents of chiral 2D conformal field theories with central charge 24 \cite{Schellekens:1992db,Hohn:2017dsm,Hohn2022}, while the type II strings correspond to currents of the chiral 2D superconformal field theory with central charge 12 given by 24 real fermions \cite{Harrison:2020wxl}. Here we present an explicit proof of this statement, which extends to non-cyclic and even to non-Abelian orbifolds. 

In fact, a classification was undertaken in \cite{Fraiman:2022aik} with the expectation of 47 moduli spaces. From these, 23 are cyclic orbifolds of both the type II and heterotic string theories and 1 is the trivial Narain moduli space of the heterotic theory.
In particular, there are 10 heterotic orbifolds in this classification. From these, 8 correspond to known constructions, with $n=2,\cdots 8$, while the three remaining theories, which are more subtle, were recently constructed in \cite{Aldazabal:2025zht}. On the other hand, the predicted type II strings add up to 13 (5 with discrete theta angle), out of which 8 (3 with discrete theta angle) are known in the literature at the moment --- more details on both heterotic and type II constructions are given below. Here we will construct explicitly the five new predicted theories. In doing so, we show that the predictions regarding discrete theta angles in \cite{Fraiman:2022aik} (which lie outside the scope of the worldsheet classification proof above) are indeed correct. Interestingly, they correspond to isolated theories, where the only massless degrees of freedom belong to the gravity multiplet and the only scalar is the dilaton. Such theories are called \textit{string islands} \cite{Harvey:1987da}.

As mentioned above for $D = 7$ we can match each of the frozen singularity backgrounds of M-theory on K3 with some perturbative string, possibly with a discrete theta angle turned on. It turns out that the effects of singularity freezing on the BPS spectrum can be determined systematically from these stringy constructions, without any reference to the K3 frame. In fact, this construction is naturally generalizable to 6D theories with $\mathcal{N} = (1,1)$ supersymmetry! \cite{Fraiman:2021hma,Fraiman:2022aik}. In this way, we can construct an abstract map that is predicted to be physically realized by frozen singularities, but more generally, relates all the $N = 16$ theories in a given dimension independently of their duality frame. 

In 6D, this abstract map predicts in a robust way the existence of 23 inequivalent string theories realized as asymmetric orbifolds of either heterotic or type II strings on $T^4$, for a total of 24 theories including the unorbifolded one.\footnote{Taking the overall mechanism at face value one can formally construct the defining data of 23 ``non-cyclic" theories, hence the expectation of a total of 47 theories in \cite{Fraiman:2022aik}, but it turns out that this particular use of the procedure is not well defined. An explicit physical realization of the rank reduction map of \cite{Fraiman:2022aik} makes it clear that its naive application in the construction of these extra theories is obstructed \cite{RR}.\label{fn1}} They may be described as follow
\begin{itemize}
    \item \underline{Heterotic:} These comprise on one hand the $S^1$ compactifications of the $\Z_n$ holonomy heterotic holonomy triples, $n = 2,3,4,5,6$ as well as $\Z_7$ and $\Z_8$ asymmetric orbifolds of the heterotic string on $T^4$ \cite{deBoer:2001wca}. On the other hand, there are three special $\Z_n$ asymmetric orbifolds, $n = 2,10,12$, exhibiting order doubling $n \to 2n$ as symmetries of the heterotic worldsheet CFTs. Explicit constructions of the latter were given recently in \cite{Aldazabal:2025zht}.\footnote{One of these theories admits two different realizations which might be inequivalent. We comment on this point in Section \ref{s:conclusions}.} Their 5D realization is simpler to obtain and can be found in \cite{Persson:2015jka}, and in 2D they are even more easily realized as holomorphically factorized CFTs. 

    \item \underline{Type II:} These are $\Z_n$ asymmetric orbifolds of type II strings on $T^4$ roughly divided into two classes. The first class is given by $n = 2,3,4,6$, and can be realized in seven dimensions, where they are dual to M theory on K3 with frozen singularities (see above); the cases $n = 2,3,4$ admit alternative versions with discrete theta angle. In six dimensions there are four new possibilities with $n = 5,8,10,12$, with the cases $n = 5,8$ admitting a nontrivial discrete theta angle. As mentioned above these theories are special since they correspond to \textit{string islands}. The $n = 5$ theory without discrete theta angle was constructed in \cite{Dabholkar:1998kv}. The 5D realizations of the four 6D string islands can be found in \cite{Persson:2015jka, Baykara:2024vss}, but their versions with discrete theta angle (for $n = 5,8$) have not been constructed in any dimension. In the absence of a discrete theta angle, all of these theories are realized as holomorphically factorized CFTs in 2D.
\end{itemize}
We emphasize that all of the above 23 theories are \textit{predicted to exist} from a formal map generalizing the frozen singularity mechanism. At the moment of writing this article, 15 theories out of the 23 had known realizations in the literature.\footnote{In \cite{Chaudhuri:1995fk}, Chaudhuri, Hockney and Lykken constructed a heterotic theory in 6D with $\mathcal{N} = (1,1)$ in the fermionic formulation which was overlooked in \cite{Fraiman:2022aik}. It likely corresponds in bosonic formulation to the $Z_2$ asymmetric orbifold of \cite{Aldazabal:2025zht}, bringing the number of previously known theories to 16. We also note that Chaudhuri-Lowe predicted in \cite{Chaudhuri:1995ee} various other 6D $\mathcal{N} = (1,1)$ theories which turn out to be excluded by our results.} The suggestion that the remaining theories admit a description in terms of asymmetric orbifolds follows from their known existence in dimensions $D \leq 5$, and is confirmed for heterotic strings by \cite{Aldazabal:2025zht}. The suggestion that two of the new type II theories have discrete theta angles turned on is a reasonable generalization from known theories of this type. 

\subsection{Summary of results and organization of the paper}

This paper is largely focused on type II asymmetric orbifolds, as the new theories that we present are of this type. Concerning these theories, we present five main results:
\begin{enumerate}
    \item We prove that every type II asymmetric orbifold in 6D $\mathcal{N} = (1,1)$, where the two gravitini come from chiral worldsheet fields (i.e. those considered in \cite{Dixon:1987yp}) is in one-to-one correspondence with an orbifold of the $E_8$ superconformal field theory, leading to a classification of such 6D theories at the perturbative level. See the \textbf{Proposition} in Section \ref{ss:classification} for the precise statement. The bulk of the proof is rather technical and we relegate it to Appendix \ref{app:class}. We also show that this proof extends to the heterotic string, and that its results are in perfect agreement with the recent explicit constructions of \cite{Aldazabal:2025zht}. 

    \item We give an explicit construction of each of the 6D theories predicted by the above classification scheme. These are $\Z_n$-orbifolds with $n = 2,3,4,5,6,8,10,12$, with $n = 8,10,12$ corresponding to \textit{new theories}.

    \item We show explicitly that the four string islands given by $Z_{5,8,10,12}$ asymmetric orbifolds, once compactified on $S^1$ to 5D, are dual to quasicrystalline compactifications constructed recently in \cite{Baykara:2024vss}. This duality relationship forms then a bridge between controlled perturbative constructions and exotic type IIA sigma models with K3 target space. 

    \item We present a general treatment of discrete theta angles in 6D $\mathcal{N} = (1,1)$ asymmetric orbifolds, and verify the prediction in \cite{Fraiman:2022aik} that those with $n = 5,8$ admit nontrivial discrete theta angles of order 5 and 2, respectively. 

    \item We show explicitly that the discrete theta angle is correlated with the presence of extra non-BPS strings extending the lattice of charges for fundamental strings according to the order of the discrete theta angle as in \cite{Montero:2022vva}. Moreover, we explicitly compute the lattices of electric charges for particles and how they are extended due to extra non-BPS states, in perfect agreement with the predicted charge lattices in \cite{Fraiman:2022aik}. 
\end{enumerate}

This paper is structured as follows. Section \ref{s:asymorb} is dedicated to the classification and construction of the new string islands as asymmetric orbifolds of type II strings without discrete theta angles, as well as their duality relations with quasicrystalline orbifolds \cite{Baykara:2024vss}. For completeness we give the explicit construction of all 8 asymmetric orbifolds. In Section \ref{s:discretetheta} we give a general treatment of discrete theta angles in asymmetric orbifolds, determining how they affect the string charge lattice as well as the particle charge lattice. We show that out of the four perturbative islands, only two admit a nontrivial discrete theta angle as expected. We discuss various aspects and consequences of our results in \ref{s:conclusions}. The bulk of the proof of the classification statement for asymmetric orbifolds is relegated to Appendix \ref{app:class}. Appendix \ref{app:theta} contains technical details regarding the computation of discrete theta angles. 

\section{Type II asymmetric orbifolds old and new}\label{s:asymorb}
In this section, we review the basic construction of asymmetric orbifolds \cite{Narain:1986qm} specialized to type II strings compactified on $T^4$ with the requirement that half of the gravitini are preserved. The resulting type II strings were of the type considered in the seminal article of Dixon, Kaplunovski and Vafa \cite{Dixon:1987yp}, and we refer the reader to that article for an in depth discussion of the worldsheet conformal field theories. We classify the set of theories of this type in six dimensions by demonstrating their correspondence with orbifolds of the 2D chiral sCFT based on the $E_8$ lattice, with the bulk of the proof relegated to Appendix \ref{app:class}, and then we proceed to construct all of them explicitly. Finally, we show that these theories are connected upon circle compactification to the 5D quasicrystallographic orbifolds, constructed recently in \cite{Baykara:2024vss}, as Sen-Vafa dual pairs \cite{Sen:1995ff}.

\subsection{Asymmetric orbifolds}\label{ss:asymorb}
Type II string compactification on $T^4$ is characterized by the even self-dual \textit{Narain lattice} $\Gamma_{4,4}\subset \mathbb R^{4,4}$, also known as the momentum-winding lattice. The T-duality group is the automorphism group of the lattice $\mathrm{Aut}(\Gamma_{4,4})\subset \mathrm{Spin}(4,4;\mathbb{Z})$.

If $\Gamma_{4,4}$ is chosen such that a T-duality $M\in \mathrm{Aut}(\Gamma_{4,4})$ decomposes into left and right blocks as
\begin{align}
M = \begin{pmatrix}
		M_L & 0 \\ 0 & M_R
	\end{pmatrix},\qquad M_L,M_R\in \mathrm{Spin}(4,\mathbb Z),
\end{align}
then it lifts to a crystalline symmetry of the worldsheet CFT.\footnote{We only consider crystalline symmetries here, therefore we use $M_L,M_R\in\mathrm{Spin}(4,\mathbb Z)$. However, for quasicrystalline symmetries \cite{Harvey:1987da,Baykara:2024vss}, it is possible that $M_L$ and $M_R$ are not conjugate to integral matrices by themselves but $M=\mathrm{diag}(M_L,M_R)$ is conjugate to an integral matrix in $\mathrm{Spin}(4,4;\mathbb Z)$.} Such T-dualities that become a symmetry form the \textit{Narain symmetry group}
\begin{align}
\mathrm{Sym}(\Gamma_{4,4}) := \mathrm{Aut}(\Gamma_{4,4})\cap (\mathrm{Spin}(4,\mathbb Z)\times \mathrm{Spin}(4,\mathbb Z)).
\end{align}
The simplest example is T-duality of the bosonic string on $S^1$ at the self-dual radius, where we have $M=\mathrm{diag}(-1,1)$.

To specify the Narain lattice, we go to a point in moduli space represented by a polarization of the Narain lattice
\begin{equation}\label{narpol}
	 W_L\oplus W_R \subset \Gamma_{4,4}\,,
\end{equation}
where $W_L$ is a signature $(4,0)$ primitive sublattice and its orthogonal complement $W_R = W_L^\perp$ is of signature $(0,4)$. Primitivity means that $(W_L^\perp)^\perp = W_L$. In all cases of interest, $W_R$ is isometric to $W_L$ up to a change in signature, and we write the Narain lattice polarized to \eqref{narpol} as $\Gamma_{4,4}(W_L)$. In all the cases we consider, $W_L$ is the root lattice of a Lie algebra $G$, and the symmetry group contains the Weyl group on the left and right
\begin{align}
W(G)_L\times W(G)_R \subset \mathrm{Sym}(\Gamma_{4,4}(G)).
\end{align}

The matrix $M_{L,R} \in Spin(4,\mathbb{Z})$ is an element connected to the identity in $Spin(4,\mathbb{R})$ and so it can be taken to lie in a maximal torus $U(1)\times U(1) \subset Spin(4)$. This means $M_{L,R}$ can be completely specified by two phases $\phi^1_{L,R}$ and $\phi^2_{L,R}$ in an appropriate basis. We arrange these two phases in a so-called \textit{twist vector}
\begin{align}
    \phi &:= (\phi_L;\phi_R),\\
	\phi_{L,R} &:= (\phi_{L,R}^1, \phi_{L,R}^2),\qquad \phi_{L,R}^{1,2}\in \mathbb Q/2\mathbb Z\,.
\end{align}
At the level of the bosonic subgroup $SO(4)$, this is nothing more than the statement that a rotation in $\mathbb{R}^4$ can be completely specified by two rotations along orthogonal 2-planes. The $\mathbb Q/2\mathbb Z$ values facilitate the spin uplift. In particular, a $2\pi$ rotation on the left or right corresponds to $(-1)^{F_{L}}$ or $(-1)^{F_R}$ respectively, denoted by the twist vector $(1,0)$. 

In addition to T-dualities, we can consider the effect of shifts along $T^4$ that act on the KK modes together with some phases on the winding modes. They are given by shift vectors in the span of the Narain lattice. We will only consider shifts of finite order, therefore we take shift vectors in the rational span of the lattice
\begin{align}
v=(v_L,v_R)\in \mathbb Q\otimes \Gamma_{4,4}. 
\end{align}
The order of the shift $v$ is the smallest integer $N>0$ such that $Nv\in\Gamma^{4,4}$.

One can \textit{orbifold} by a roto-translation $g=(M,v)$ of order $N$ by gauging it on the worldsheet. In particular, orbifolding consists of relaxing the boundary conditions as
\begin{align}
    X_{L,R}(\sigma+2\pi,\tau) = g_{L,R}^n\cdot X_{L,R}(\sigma,\tau),\qquad n\in \mathbb Z_N
\end{align}
where $g_{L}=(M_L,v_L)$, $g_R=(M_R,v_R)$, and $X_{L,R}$ are left and right mover worldsheet fields, and projecting to the invariant states
\begin{align}
\hat{g}\cdot \ket{\psi} = \ket{\psi},
\end{align}
where $\hat{g}$ is the lift of $g$ to the worldsheet CFT. The sector whose boundary condition is changed by $g^n$ is called the $n^{\mathrm{th}}$ \textit{twisted sector}. We specialize to \textit{asymmetric orbifolds} where the left and right actions are not the same $M_L\neq M_R$. 

In our case, we aim to preserve 16 of the original 32 supercharges. Therefore, we choose our orbifolding action $g=(M,v)$ with a trivial rotation on the right movers $M_R=\mathbb{1}$
\begin{equation}\label{trivR}
	M = \begin{pmatrix}
		M_L & 0 \\ 0 & \mathbb{1}
	\end{pmatrix} \in Spin(4;\mathbb{Z})\times Spin(4;\mathbb{Z}) \subset Spin(4,4;\mathbb{Z}).
\end{equation}
This ensures that all the NSR gravitini are preserved and all the RNS gravitini are projected out, which implies that the number of supercharges is halved. The shift vector $v$ ensures that there is no accidental supersymmetry enhancement to $\mathcal{N} = (2,2)$ in the twisted sectors.

\subsection{Classification}\label{ss:classification}
We are interested in classifying asymmetric orbifolds of type II strings on $T^4$ preserving 16 supercharges. In principle this class of theories could contain orbifolds of non-cyclic type, or more generally, non-Abelian type, but it turns out that \textit{cyclic orbifolds are exhaustive}, as we will describe below. The relevant ingredients are the symmetries $M$ and shifts $v$ discussed in the previous section, and the classification problem can be separated into two steps:
\begin{enumerate}
    \item Classify every T-duality symmetry $M$ of the type II string on $T^4$ preserving the worldsheet superalgebra, i.e. leading to a consistent type II string orbifold.

    \item For each $M$, classify the possible shift vectors $v$ such that 16 supercharges are preserved when $g = (M,v)$ is gauged in the worldsheet. 
\end{enumerate}
The classification program in step one was carried out by Volpato in \cite{Volpato:2014zla}. It then remains to determine if these are compatible with the inclusion of nontrivial shift vectors along the invariant lattice. 

To solve this problem it helps to work with the following setup. Consider the type II string on $T^4$ and compactify it further on another $T^4$. The Narain lattice is now $\Gamma_{8,8}$, which at a special point in moduli space is orthogonally split as
\begin{equation}
    \Gamma_{8,8} \simeq E_8 \oplus E_8(-1)\,,
\end{equation}
where $E_8(-1)$ denotes the $E_8$ lattice with a change of sign in its bilinear form. At this point, the worldsheet CFT is factorized into two copies of the $E_8$ sCFT, one holomorphic and one antiholomorphic. At the level of lattice automorphisms, the symmetries $M$ of the original $T^4$ compactification turn out to be in correspondence with even Weyl symmetries $M'$ of the $E_8$ root lattice $W^+(E_8)$ which fix a sublattice of rank at least 4 \cite{Volpato:2014zla}. This relationship uplifts to one of CFT symmetries in the $T^4\times T^4$ compactification as the statement that $M$ and $M'$ are conjugate under the full T-duality group. In other words, a 6D asymmetric orbifold by $g = (M,0)$ is equivalent to a 2D ``holomorphic orbifold" by $g' = (M',0)$.

Now we ask if the correspondence between $M$ and $M'$ can be extended to a correspondence between $g = (M,v)$ and $g = (M',v')$ with nontrivial shift vectors $v$ and $v'$, where $v'$ has no components along the $E_8(-1)$ lattice. In other words, we ask if there is a correspondence at the level of the full symmetry group. If this is the case, cyclic orbifolds of the 6D theory will be in one to one correspondence with cyclic orbifolds of the $E_8$ sCFT. We may ask moreover if this correspondence extends to $n$-tuples of symmetries $g_1,...,g_n$ with or without nontrivial shift vectors, in which case the following statement holds:\\

\noindent\fbox{
\parbox{\textwidth}{%
\vspace{0.1in}
\textbf{Proposition:} Any asymmetric orbifold of the type II string on $T^4$ by a symmetry group $\mathcal{G}$ preserving the $\mathcal{N}=(1,1)$ structure and acting trivially on right (or left) moving gravitini is in correspondence with an orbifold of the $E_8$ sCFT by some symmetry group $\mathcal{G}'$ which acts trivially on a sublattice of rank at least 4 and preserves the $\mathcal{N} = 1$ structure. This correspondence is realized as an equivalence between the $T^4$ compactified type II orbifold and the tensor product of the $E_8$ sCFT orbifold with an antiholomorphic $E_8$ sCFT.
\vspace{0.1in}
}
}\\

\noindent This proposition turns out to be true, and we present a proof in Appendix \ref{app:class}. As a consequence it suffices to classify the nontrivial orbifolds of the $E_8$ sCFT of the stated form.\\

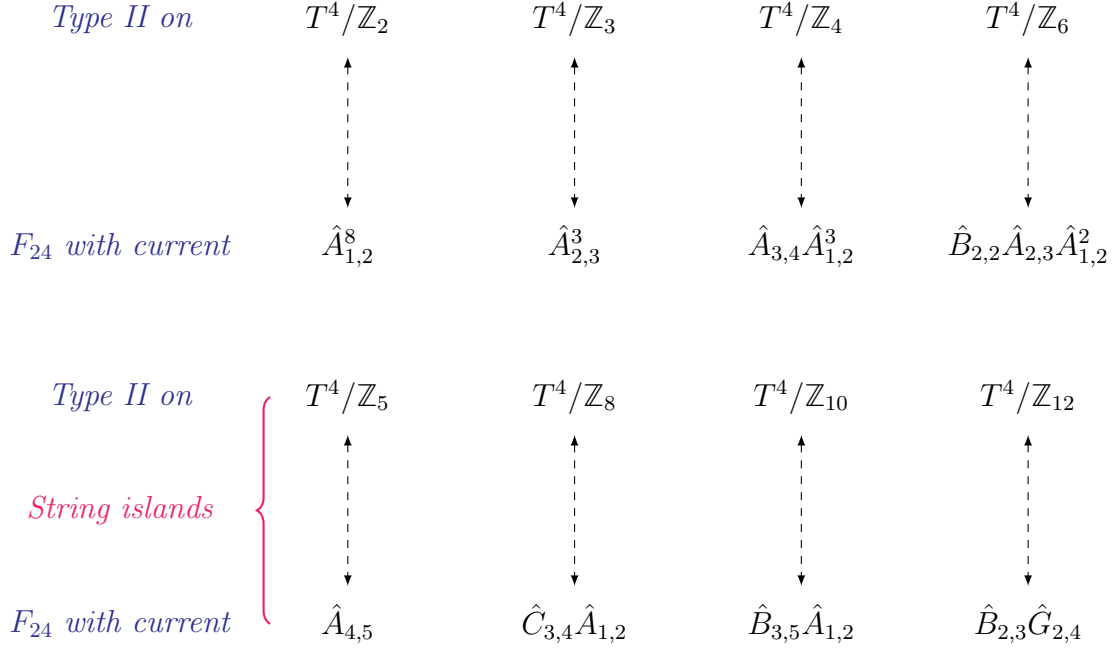
\begin{figure}
    \centering
    \begin{tikzpicture}
    \draw[latex-latex,dashed,](3,2.5)--(3,0.5);
    \draw[latex-latex,dashed,](6,2.5)--(6,0.5);
    \draw[latex-latex,dashed,](9,2.5)--(9,0.5);
    \draw[latex-latex,dashed,](12,2.5)--(12,0.5);
    \draw(0,3)node{\textcolor{Blue}{\textit{Type II on}}};
    \draw(3,3)node{$T^4/\mathbb{Z}_2$};
    \draw(6,3)node{$T^4/\mathbb{Z}_3$};
    \draw(9,3)node{$T^4/\mathbb{Z}_4$};
    \draw(12,3)node{$T^4/\mathbb{Z}_6$};
    
    \draw(0,0)node{\textcolor{Blue}{$F_{24}$ \textit{with current}}};
    \draw(3,0)node{$\hat{A}_{1,2}^8$};
    \draw(6,0)node{$\hat{A}_{2,3}^3$};
    \draw(9,0)node{$\hat{A}_{3,4}\hat{A}_{1,2}^3$};
    \draw(12,0)node{$\hat{B}_{2,2}\hat{A}_{2,3}\hat{A}_{1,2}^2$};
    \begin{scope}[shift={(0,-5)}]
    \draw(0,3)node{\textcolor{Blue}{\textit{Type II on}}};
    \draw(0,0)node{\textcolor{Blue}{$F_{24}$ \textit{with current}}};
    \draw[latex-latex,dashed,](3,2.5)--(3,0.5);
    \draw[latex-latex,dashed,](6,2.5)--(6,0.5);
    \draw[latex-latex,dashed,](9,2.5)--(9,0.5);
    \draw[latex-latex,dashed,](12,2.5)--(12,0.5);
    \draw [WildStrawberry,thick,decorate,decoration={brace,amplitude=5pt,mirror,raise=4ex}]
  (2.7,3) -- (2.7,0);
  \draw(0,1.5)node{\textcolor{WildStrawberry}{\textit{String islands}}};
    \draw(3,3)node{$T^4/\mathbb{Z}_5$};
    \draw(6,3)node{$T^4/\mathbb{Z}_8$};
    \draw(9,3)node{$T^4/\mathbb{Z}_{10}$};
    \draw(12,3)node{$T^4/\mathbb{Z}_{12}$};
    
    \draw(3,0)node{$\hat{A}_{4,5}$};
    \draw(6,0)node{$\hat{C}_{3,4}\hat{A}_{1,2}$};
    \draw(9,0)node{$\hat{B}_{3,5}\hat{A}_{1,2}$};
    \draw(12,0)node{$\hat{B}_{2,3}\hat{G}_{2,4}$};
    \end{scope}
    \end{tikzpicture}
    \caption{Correspondence between asymmetric orbifolds of the type II string on $T^4$ and the 2D superconformal field theory $F_{24}$ comprising 24 fermions and a current specified by the given Kac-Moody algebra. The rank of the type II theory is equal to the rank of the corresponding current minus 4, hence the theories in the lower row have rank 0 and are accordingly referred to as
    string islands following \protect\cite{Dabholkar:1998kv}. The symbol $\hat{A}_{m,n}$ denotes the Kac-Moody algebra for $SU(m+1)$ at level $n$, and likewise, $\hat{B}_{n,m}$, $\hat{C}_{m,n}$ and $\hat{G}_{2,m}$ refer to $SO(2m+1)$, $Sp(m)$ and $G_2$.}     \label{fig:corresp}
\end{figure}

The requirement that the $E_8$ sCFT orbifold preserves the $\mathcal{N} = 1$ structure constrains the resulting theory to be another holomorphic sCFT with central charge 12. There exist three such theories, namely (1) the $E_8$ sCFT itself, (2) the sCFT $F_{24}$ comprising 24 free chiral fermions and (3) the sCFT analog of the monster CFT (see \cite{Harrison:2020wxl} and references therein). The first case is trivial, and arises in particular when $v = 0$, while the third case is given by an action without rank 4 invariant lattice. The only suitable nontrivial orbifold is then given by $F_{24}$, which comes in eight different variants depending on its current algebra structure, see Figure \ref{fig:corresp}. It follows that there exist only eight inequivalent asymmetric orbifolds of the type II string on $T^4$ preserving 16 supercharges, and these correspond to the theories predicted in \cite{Fraiman:2022aik} as discussed in the Introduction. We now go through their explicit construction.
\subsection{Explicit constructions}\label{ss:explicit}
Among the eight asymmetric orbifolds, five are already known in various forms in the literature. To keep our discussion clear and complete, however, we will record their construction along with the new theories. These are naturally split into two classes, depending on how the orbifold operation $g$ acts on the lattice of RR 1-form field charges $\Gamma_{4,4}^-$:\\
\begin{enumerate}
    \item \underline{RR-Crystalline action:} The lattice automorphism $M$ acts symmetrically on the lattice $\Gamma_{4,4}^-$ such that the equivalent action on the Narain lattice $\Gamma_{4,4}$ defines a geometric K3 sigma model. These orbifolds are
    \begin{equation}
        \Z_2\,, ~~~~~\Z_3\,, ~~~~~\Z_4\,, ~~~~~\Z_6\,.
    \end{equation}
    They are dual to M-theory on $(T^4\times S^1)/\mathbb{Z}_n$ where the orbifold action on the $T^4$ coordinates is as above, together with a shift along $S^1$. These latter compactifications admit an uplift to F-theory on $(T^4\times S^1)/\mathbb{Z}_n$ \cite{deBoer:2001wca} or equivalently type IIB on $(T^2\times S^1)/\mathbb{Z}_n$ with a nontrivial action of the metaplectic cover of $SL(2,\Z)$ on the axiodilaton \cite{Montero:2022vva}.

    \item \underline{RR-Quasicrystalline action:} The lattice automorphism $M$ acts asymmetrically on $\Gamma_{4,4}^-$ in a way that does not descend to separate symmetries on its negative and positive definite sublattices, i.e. it acts in a quasicrystalline fashion \cite{Harvey:1987da,Baykara:2024vss}. These orbifolds are
    \begin{equation}
        \Z_5\,, ~~~~~\Z_8\,, ~~~~~\Z_{10}\,, ~~~~~\Z_{12}\,.
    \end{equation}
    They are characterized by a lack of Narain moduli, and their dual frames at strong coupling are so far unknown.
\end{enumerate}
We will come back to the relationship between K3 sigma models and asymmetric orbifolds in Section \ref{ss:quasicrystals}, which can be understood in terms of a certain duality between five dimensional theories in the Sen-Vafa framework \cite{Sen:1995ff}. In the following we will record the data specifying the construction of each of the asymmetric orbifolds. 

\subsubsection{The four conventional orbifolds}\label{sss:conv}
We start with the four conventional orbifolds given by RR-crystalline actions.\\

\noindent \underline{$\Z_2$ asymmetric orbifold:} The symmetry $M$ acts as the identity on the Narain lattice. Its CFT action is $(-1)^{F_L}$ where $F_L$ is the left-moving worldsheet fermion number. Its corresponding twist vector reads
\begin{equation}
    \phi_L = (1,0)\,.
\end{equation}
The shift vector $v$ can be chosen as the geometric order 2 translation along one of the 1-cycles in the $T^4$, in which case the asymmetric orbifold implements a $(-1)^{F_L}$ Wilson line. This theory is then a $T^3$ compactification of a 9D $\mathcal{N} = 1$ asymmetric orbifold. There are two such inequivalent theories depending on the parent theory being type IIA or type IIB. The former was studied in detail in \cite{Hellerman:2005ja}, and a comprehensive analysis of both theories and their duality frames at the level of their moduli spaces was carried out in \cite{Aharony:2007du}. Upon $S^1$ compactification the two 9D theories become T-dual, and so the 6D asymmetric orbifold is unique up to worldsheet marginal deformations.\\

\noindent \underline{$\Z_3$ asymmetric orbifold:} The symmetry $M$ acts on the Narain lattice by a $\Z_3$ Weyl rotation of an $A_2$ sublattice, and has corresponding twist vector
\begin{equation}
    \phi_L = (2/3,0)\,.
\end{equation}
The appropriate Narain lattice polarization is 
\begin{equation}
    \Gamma_{4,4} \simeq \Gamma_{2,2}(A_2)\oplus \Gamma_{2,2}\,,
\end{equation}
and we may take $v$ to generate the geometric order 3 translation along one of the cycles associated to the invariant $\Gamma_{2,2}$. This is a simple example of a flat T-fold or monodrofold, since the asymmetric orbifold construction implements a T-duality monodromy/Wilson line (see e.g. \cite{Hellerman:2006tx} and references therein). It is a circle compactification of a 7D theory of rank 1.\\

\noindent \underline{$\Z_4$ asymmetric orbifold:} The symmetry $M$ acts on the Narain lattice by a $\Z_2$ reflection of a $2\,A_1$ sublattice, and may be taken to act on the complex structure $\tau$ and complexified Kähler structure $\rho$ of a $T^2$ as
\begin{equation}\label{fullT}
    M: ~~~ \tau \to -\frac{1}{\tau}\,, ~~~~~\rho \to -\frac{1}{\rho}\,.
\end{equation}
This operation is the full T-duality on $T^2$ which generalizes $R \to 1/R$, and is a priori a T-duality \textit{symmetry} of the type II string. However, as explained in \cite{Hellerman:2006tx}, a naive gauging of this symmetry is inconsistent with level-matching of the spectrum. Indeed, in accordance with the results of \cite{Volpato:2014zla}, where the corresponding conjugacy class is denoted $2B$, the worldsheet CFT action is necessarily an order 4 lift of the lattice automorphism (equivalently the T-duality \ref{fullT}) that squares to $(-1)^{F_L}$. This is the lift from orthogonal matrices in $SO(4,\Z)$ to $Spin(4,\Z)$ in eq. \eqref{trivR}, and is accordingly encoded in a twist vector 
\begin{equation}
    \phi_L = (1/2,0)\,.
\end{equation}
The Narain lattice polarization is
\begin{equation}
    \Gamma_{4,4} \simeq \Gamma_{2,2}(2\,A_1)\oplus \Gamma_{2,2}\,,
\end{equation}
and as before, we may take $v$ to generate the geometric order 4 translation along one of the cycles associated to the invariant $\Gamma_{2,2}$. This theory is also a circle compactification of a 7D theory of rank 1. 

\noindent \underline{$\Z_6$ asymmetric orbifold:} This theory is constructed in the same manner as the $\Z_3$ orbifold, with the difference that the $Spin(4,\Z)$ lift of the lattice automorphism is taken to be of order 6, i.e. such that its third power is $(-1)^{F_L}$. Indeed the existence of this orbifold is implied by the existence of the $\Z_2$ and the $\Z_3$ orbifold above given that $\Z_6 = \Z_2\times \Z_3$. This is the third and last theory arising as a compactification of a 7D theory of rank 1. The twist vector is
\begin{equation}
    \phi_L = (1/3,0)\,,
\end{equation}
in accordance with the relation $(1,0)+(2/3,0) = (1/3,0)$ as half-phases up to an overall sign.

\subsubsection{String islands}
Now we construct the four string islands given by asymmetric orbifolds with RR-quasicrystalline actions. The $\Z_5$-orbifold appeared first in \cite{Dabholkar:1998kv} while the other three are new. Note that the shift vectors are presented in the alpha basis.\\

\noindent \underline{$\Z_5$ string island:} The symmetry $M$ acts as the order 5 Weyl rotation of $A_4$ at the point in moduli space with Narain lattice $\Gamma_{4,4}(A_4)$. The twist vector is
\begin{equation}
    \phi_L = (2/5,4/5)\,.
\end{equation}
Note that both $U(1)$ phases are now nontrivial, and there are no spectator cycles. The shift vector $v$ must then be chosen to lie in the rational span of $A_4(-1)$ which is a negative definite invariant lattice. There turns out to be a suitable choice
\begin{equation}
    v_L = \frac 1 5(2,-2,0,4,-4)\,.
\end{equation}
This is the asymmetric orbifold with largest prime order, and as we will see in Section \ref{s:discretetheta}, it admits an order five discrete theta angle leading to the strongest violation of BPS completeness for the string charge lattice in 6D $\mathcal{N} = (1,1)$ theories.\\

\noindent \underline{$\Z_8$ string island:} The symmetry $M$ acts on a sublattice $A_3A_1\subset \Gamma_{4,4}$ by a reflection on $A_1$ and the order four Weyl rotation of $A_3$. Similarly to the $\Z_4$ asymmetric orbifold, the order of $M$ is doubled in $Spin(4,\Z)$, and the twist vector is
\begin{equation}
    \phi_L = (1/4,1/2)\,.
\end{equation}
Note that the $\Z_2$ subgroup generated by $M^4$ has twist vector $4\,\phi_L = (1,2) \simeq (1,0)$ in reflecting that $M^4 = (-1)^{F_L}$. The $\Z_4$ subgroup generated by $M^2$ has in turn a twist vector $2\,\phi_R = (1/2,1)$, and so $M^2$ is the product of the symmetry defining the $\Z_4$ asymmetric orbifold and $(-1)^{F_L}$. This is the alternative but equivalent lift of the lattice automorphism reflecting $2A_1$ to a worldsheet symmetry \cite{Volpato:2014zla}. The shift vector can be taken as
\begin{equation}
    v_L = \frac 1 8(2,0,-3,1,1,-1)
\end{equation}
in the negative definite part of $\Gamma_{4,4}(A_3A_1)$. As we will see, like the $\Z_2$ and $\Z_4$ orbifolds to which this theory is related, it admits a $\Z_2$ discrete theta angle.\\

\noindent \underline{$\Z_{10}$ string island:} As with the $\Z_6$ asymmetric orbifold, consistency requires the existence of a $\Z_{10} = \Z_2 \times \Z_5$ string island, corresponding to a second uplift of the lattice automorphism rotating $A_4$ with doubled order in $Spin(4)$. The twist and shift vectors are
\begin{align}
    \phi_L &= (1/5,2/5),\\
    v_L &= \frac 1 {10}(0,1,-1,1,-1),\\
    v_R &= \frac 1 {10}(-3,5,1,-6,3).
\end{align}
Note that the shift on the right $v_R$ affects only the $5$th twisted sector where the twist is $(-1)^{F_R}$.
\\ 

\noindent \underline{$\Z_{12}$ string island:} Since the $\Z_3$ and $\Z_4$ orbifolds act respectively as automorphisms on the rank 2 sublattices $A_2$ and $2A_1$ of $\Gamma_{4,4}(A_2A_1A_1)$, and $\Z_{12} = \Z_3 \times \Z_4$, combining these actions on the sublattice $A_2+2A_1$ yields a string island provided a suitable shift exists. This turns out to be the case, and the twist and shift vectors are
\begin{align}
    \phi_L &= (1/3,1/2),\\
    v_L &= \frac 1 {12} (1,2,-3,1,-1,2,-2).
\end{align}

We summarize the data of the construction of these string islands in Table \ref{tab:islands}, where we also include the order of their allowed discrete theta angles. We will explain in detail how these discrete theta angles arise and what are their effects in Section \ref{s:discretetheta}.

\begin{table}[h!]
\centering
\begin{tabular}{|c|c|c|c|}
\hline
& & & \\
Islands   & Lattice   & Action & Theta angle \\ & & & \\ \hline\hline & & & \\
$\Z_5$    & $\Gamma^{4,4}(A_4)$    & \begin{tabular}[c]{@{}c@{}}$\phi_L = (2/5,4/5)$\\ $v_R = \frac 1 5(2,-2,0,4,-4)$\end{tabular}                      &             $\Z_5$\\ & & & \\ \hline & & & \\
$\Z_8$    & $\Gamma^{4,4}(A_3A_1)$ & \begin{tabular}[c]{@{}c@{}}$\phi_L = (1/4,1/2)$\\ $v_R = \frac 1 8(2,0,-3,1,1,-1)$\end{tabular}                      &             $\Z_2$\\ & & & \\\hline & & & \\
$\Z_{10}$ & $\Gamma^{4,4}(A_4)$    & \begin{tabular}[c]{@{}c@{}}$\phi_L = (1/5,2/5)$\\ $v_L= \frac 1 {10}(-3,5,1,-6,3)$\\ $v_R=\frac 1 {10}(0,1,-1,1,-1)$\end{tabular} &            $1$ \\ & & & \\ \hline & & & \\
$\Z_{12}$ & $\Gamma^{4,4}(A_2A_1A_1)$    & \begin{tabular}[c]{@{}c@{}}$\phi_L = (1/3,1/2)$\\ $v_R = \frac 1 {12} (1,2,-3,1,-1,2,-2)$\end{tabular}                     &             $1$\\ & & & \\ \hline
\end{tabular}
\caption{This table summarizes the four string islands $\Z_{5,8,10,12}$ without discrete theta angle. Note that the $\Z_5$ island first appeared in \protect\cite{Dabholkar:1998kv}. The last column gives the order of the allowed discrete theta angle, non-trivial only for the $\Z_{5,8}$ string islands.}
\label{tab:islands}
\end{table}

\subsection{Duality with quasicrystalline orbifolds}\label{ss:quasicrystals}
As we have alluded to, the orbifolds defining the string islands have a quasicrystalline action on the lattice of RR 1-form charges. We will now explain in more detail why this is the case, and use this fact to demonstrate a non-trivial duality involving these theories. To this end we will use a formalism of Sen and Vafa \cite{Sen:1995ff} for constructing dual pairs of freely acting type II string orbifolds in $D = 5,6$ using U-duality and the adiabatic argument. 

The essential idea of Sen and Vafa is illustrated by considering two different orbifolds of the type II string on $T^4\times S^1$ given by T-dualities $M, M' \in SO(4,4,\Z)$ acting on the $T^4$ moduli as well as a shift along $S^1$.
The full non-perturbative U-duality group of the theory is $SO(5,5,\Z)$ and we may choose $M$ and $M'$ such that they are conjugate under U-duality \textit{but not} under T-duality. The two orbifolds are then perturbatively inequivalent, yet related by some non-perturbative string duality. This latter equivalence is due to the presence of the shift, allowing to use adiabatic argument of \cite{Vafa:1995gm}.  

In the class of models of dual pairs considered in \cite{Sen:1995ff}, conjugation is given by an element $\bar{\Omega}_0 \in SO(5,5,\Z)$ inducing a transformation
\begin{equation}
    \Phi \mapsto -\Phi \,, ~~~~~ H_{\mu\nu\lambda} \mapsto e^{-2\Phi}\widetilde{H}_{\mu\nu\lambda}\,,
\end{equation}
where $\Phi$ is the 6D dilaton, $H_{\mu\nu\lambda}$ the field strength of the NSNS 2-form and $\widetilde{H}_{\mu\nu\lambda}$ its dual tensor. Conjugation by $\bar \Omega_0$ therefore relates two 5D theories through a form of 6D electromagnetic string-string duality. 

Conjugation by $\bar \Omega_0$ also relates in a very elegant way the actions of $M$ and $M'$ on the NSNS and RR 1-forms fields. Namely, \textit{it exchanges them}. This means in particular that the action on the Narain lattice $\Gamma_{4,4}$ and the lattice of RR 1-form charges $\Gamma_{4,4}^-$ is traded under this string-string duality. The elements in these lattices transform respectively as vector and cospinor representations of $Spin(4,4,\Z)$ (following the conventions of \cite{Sen:1995ff}), and triality allows in turn to write the spinorial representation of $M$ acting on $\Gamma_{4,4}^-$ as an element of $SO(4,4,\Z)$. 

In terms of twist vectors, the cospinor representation $\phi_c$ is related to the vector representation $\phi$ through the formula
\begin{align}
\begin{pmatrix}
\phi_{L,c}^1\\
\phi_{L,c}^2\\
\phi_{R,c}^1\\
\phi_{R,c}^2
\end{pmatrix} &= \frac 1 2\begin{pmatrix}
1 & -1 & 1 & -1\\
-1 & 1 & 1 & -1\\
1 & 1 & 1 & 1\\
-1 & -1 & 1 & 1
\end{pmatrix}\begin{pmatrix}\phi_L^1\\
\phi_L^2\\
\phi_R^1\\
\phi_R^2
\end{pmatrix}\,.
\end{align}
We record the twist vectors $\phi$ defining the string islands, cf. Table \ref{tab:islands}, and their associated twists $\phi_c$ in Table \ref{tab:island-quasi}. Now consider the 5D theories defined by a twist $\phi$ on the $T^4$ Narain lattice given by $\phi_c$ above, together with a shift with appropriate order along the extra $S^1$. The four theories obtained in this way are precisely the 5D quasicrystalline compactifications of \cite{Baykara:2024vss}. The construction of Sen-Vafa tells us then that taking the 6D dilaton $\Phi$ to large values we obtain the four string islands discussed in this text, compactified on $S^1$ and with a change in the underlying shift vector from the $T^4$ lattice to the $S^1$ lattice. 

\begin{table}[h!]
\centering
\begin{equation*}
\begin{array}{|c|c|c|}
\hline
\text{Islands}&(\phi_L;\phi_R) & (\phi_{L,c};\phi_{R,c})
\\\hline
\mathbb Z_5 &(2/5,4/5;0,0) & (-1/5,-1/5;3/5,3/5)\\
\mathbb Z_8 &(1/4,1/2;0,0) & (-1/8,-1/8;3/8,3/8)\\
\mathbb Z_{10} &(1/5,2/5;0,0) & (-1/10,-1/10;3/10,3/10)\\
\mathbb Z_{12} &(1/3,1/2;0,0) & (-1/12,-1/12;5/12,5/12)\\
\hline
\end{array}
\end{equation*}
\caption{Actions $(\phi_L,\phi_R)$ for the string islands correspond to the quasicrystalline actions $(\phi_{L,c},\phi_{R,c})$ in 5d after putting them on a circle and doing a duality transformation by $\bar\Omega_0$.}
\label{tab:island-quasi}
\end{table}

If we do not include a shift in the quasicrystal construction, we end up with a special type of type IIA K3 sigma model \cite{Baykara:2024vss}, which is intrinsically stringy. These 6D exotic theories are obtained from the 5D quasicrystals by taking the limit of small radius of the $S^1$. This implies that the string islands compactified on $S^1$ are continuously connected with these exotic K3 sigma models. Let us also note that the Narain lattice of the parent theory of string islands and quasicrystalline orbifolds are vastly different. Most strikingly, string islands are rational CFTs whereas quasicrystalline are irrational \cite{Harvey:1987da}.

Let us note for completeness that the 0-form charges as well as the 2-form charges transform in the spinorial representation given by a twist $\phi_s$ as
\begin{align}
\begin{pmatrix}
\phi_{L,s}^1\\
\phi_{L,s}^2\\
\phi_{R,s}^1\\
\phi_{R,s}^2
\end{pmatrix} &= \frac 1 2\begin{pmatrix}
1 & 1 & 1 & -1\\
1 & 1 & -1 & 1\\
1 & -1 & 1 & 1\\
-1 & 1 & 1 & 1
\end{pmatrix}\begin{pmatrix}\phi_L^1\\
\phi_L^2\\
\phi_R^1\\
\phi_R^2
\end{pmatrix}\,.
\end{align}
Conjugation by $\bar{\Omega}_0$ leaves invariant this action. Indeed, the difference between the dual theories lies in the origin of their gauge bosons, but in both cases the extra 0-forms and 2-forms are completely projected out.

\section{Discrete theta angles}\label{s:discretetheta}
In this section we work out in generality how discrete theta angles can appear in the asymmetric orbifolds considered in this paper and how these affect the spectrum of electric charges for strings, making it BPS-incomplete. We show that these theta angles can be realized only in five of the asymmetric orbifolds considered in Section 2. The case of the $\mathbb{Z}_2$ asymmetric orbifold was worked out in \cite{Montero:2020icj}. An analysis encompassing the $\mathbb{Z}_{2,3,4,6}$ was also carried out in this reference in a dual frame given by type F-theory compactified on $(K3\times S^1)/\mathbb{Z}_{2,3,4,6}$, with the cases $\mathbb{Z}_{2,3,4}$ admitting non-trivial discrete theta angles. We recover these results as particular cases of our analysis in the asymmetric orbifold frame. The results concerning the islands, given by $\mathbb{Z}_{5,8,10,12}$ orbifolds are new, and we show that, as predicted in \cite{Fraiman:2021hma}, only the cases $\mathbb{Z}_{5,8}$ admit discrete theta angles.  

We will also show explicitly how the discrete theta angle is correlated with the existence of non-BPS particles extending the corresponding charge lattice, just as for strings. The resulting lattices are seen to match exactly those coming from naive lattice procedures in M-theory on a K3 surface with frozen singularities \cite{ParraDeFreitas:2022wnz}, as well and the more general rank reduction map of \cite{Fraiman:2022aik}. 

\subsection{The AOB background}\label{ss:AOB}
For illustration purposes we start by reviewing how the discrete theta angle is realized in the $\Z_2$ asymmetric orbifold of the type IIB string (AOB) in nine dimensions \cite{Montero:2022vva}. The AOB is realized by a $(-1)^{F_L}$ holonomy along the internal $S^1$, cf. section \ref{sss:conv}. This holonomy acts on the RR sector by a change of sign, projecting out all of the RR $p$-form fields at the locus in moduli space with vanishing axion $C_0$. At $C_0 = 1/2$ there exists another symmetry $(-1)^{F_L}T$ where $T$ is the element of $SL(2,\mathbb{Z})$ acting as a translation $C_0 \to C_0 + 1$. Turning on a holonomy for this U-duality symmetry leads to a theory which is perturbatively equivalent to the AOB, but differs at the level of non-perturbative states. The difference between both theories is measured by the presence of a nontrivial discrete theta angle $C_0 = 1/2$ in the latter. 

To see how the spectrum of string changes, recall that $T$ acts also on the NSNS and RR 2-form fields:
\begin{equation}
    T: ~~~ \begin{pmatrix}C_2 \\ B_2\end{pmatrix} \mapsto \begin{pmatrix}C_2 + B_2 \\ B_2\end{pmatrix} ~~~\Rightarrow ~~~  (-1)^{F_L}T: ~~~ \begin{pmatrix}C_2 \\ B_2\end{pmatrix} \mapsto \begin{pmatrix}-C_2 - B_2 \\ B_2\end{pmatrix}\,.
\end{equation}
It follows that, in this construction, it is the combination $B_2+2C_2$ on which the orbifold symmetry acts as $-1$, and not $C_2$. The coupling of a $(p,q)$ string to the 2-form fields can be rewritten as
\begin{equation}
    \int qB_2 + pC_2 = \int \left(q-\frac{p}{2}\right)B_2 + \frac{p}{2}(B_2+2C_2)\,,
\end{equation}
hence a D1-brane acquires half an unit of $B_2$-field charge, extending the corresponding charge lattice from $\Z$ to $\tfrac12 \Z$. This string is also charged with respect to the 2-torsional 2-form field $B_2+2C_2$, hence it is not BPS. The string charge lattice is thus BPS incomplete. 

We now extend the analysis of \cite{Montero:2022vva} to the lattice of electric charges for particles. In the AOB, the holonomy gives rise to states with half-integral winding in the twisted sector, and so the charge lattice is extended from the Narain lattice $\Gamma_{1,1}$ to $\Gamma_{1,1}(\tfrac12)$. Non-BPS strings have half-integral charge with respect to the B-field, giving rise non-BPS particles with integral winding which have effectively the same charge as the twisted states above. The overall electric charge lattice is hence equivalent to that of the original AOB. Compactifying on an extra circle leads to different lattices, however. There are no states with half-integral winding along the extra circle, but the non-BPS strings wrap it extending the extra $\Gamma_{1,1}$ to $\Gamma_{1,1}(\tfrac12)$. For a $T^n$ compactification down to $9-n$ spacetime dimensions, we have then
\begin{equation}
\begin{split}
    \Gamma_\text{AOB}\left.\right|_{C_0 = 0} &= \Gamma_{1,1}(\tfrac12)\oplus \Gamma_{n,n}\,,\\
    \Gamma_\text{AOB}\left.\right|_{C_0 = \tfrac12} &= \Gamma_{1,1}(\tfrac12)\oplus \Gamma_{n,n}(\tfrac12)\,.\\
\end{split}
\end{equation}
Note that these lattices account only for stringy states and not other nonperturbative charges appearing at sufficiently low number of dimensions. 

We may also choose to normalize the fundamental B-field charge to 1 instead of $1/2$, so that we have instead
\begin{equation}
    \Gamma_\text{AOB}\left.\right|_{C_0 = \tfrac12} = \Gamma_{1,1}\oplus \Gamma_{n,n}\,.
\end{equation}
This normalization is more agreeable to a comparison with results coming from the frozen singularity frame. In eight dimensions, for example, the ADE type of a frozen singularity in F-theory on an elliptic K3 is $D_8$. The AOB is dual to a configuration with two frozen singularities, and the discrete theta angle is seen in the torsional part of the Mordell-Weil group of the elliptic fibration. With the discrete theta angle turned off, the middle cohomology sublattice corresponding to the pair of frozen singularities is $(D_8\oplus D_8)^+$ where the $+$ means adding the vector in the $(v,v)$ conjugacy class of $D_8\oplus D_8$. Turning on the discrete theta angle, the frozen singularity lattice is instead $D_8\oplus D_8$. Taking the orthogonal complement of each lattice within $\Gamma_{2,18}$ one obtains respectively $\Gamma_{1,1}\oplus \Gamma_{1,1}(2)$ and $\Gamma_{2,2}(2)$, coinciding with our results up to an overall normalization by a factor of 2 \cite{ParraDeFreitas:2022wnz}. This procedure of taking orthogonal complements of lattices associated to singularities coincides with the formal construction of rank reduction maps in \cite{Fraiman:2022aik} down to six dimensions, and agrees with the results above. 

\subsection{The general case}
Let us now set up the general case. In the type IIA string compactified on $T^4$, the RR axions are given by the components of the 10D 1-form and 3-form fields along the corresponding 1-cycles and 3-cycles of the $T^4$. We will label these as $C_i$ and $C_{ijk}$, $i,j,k = 1,2,3,4$. These fields can be arranged into an 8-tuple
\begin{equation}
    \psi = (C_1,C_2,C_3,C_4,C_{234},C_{134},C_{124},C_{123})
\end{equation}
which transforms as a spinor of the T-duality group $Spin(4,4;\mathbb{Z})$. The RR axions are normalized such that their shift symmetries are integer. We write the components of $\psi$ as $\psi_\alpha$, with $\alpha = (a,\bar a)$, $a = 1,2,3,4$.

In this basis, the action of elements of $Spin(4,4;\mathbb{Z})$ on $\mathcal{\psi}$ are represented by $8\times 8$ matrices $S_{\alpha}{}^\beta$ with integer entries preserving the metric of the even self-dual lattice $\Gamma_{4,4}$ written as
\begin{equation}
    \eta = \begin{pmatrix}0_4 & 1_4 \\ 1_4 & 0_4 \end{pmatrix}\,.
\end{equation}
We may choose our orbifold action to act independently along the first and last four components of $\psi$, i.e. such that the $C_i$ and $C_{ijk}$ do not mix. This condition restricts $S_\alpha{}^\beta$ to take the form
\begin{equation}\label{S}
    S = \begin{pmatrix}S' & 0 \\ 0 & (S'^{-1})^T\end{pmatrix}
\end{equation}
with $S', (S'^{-1})^T \in GL(4,\mathbb{Z})$. The RR fields transform then as
\begin{equation}\label{Saction}
    \psi_a \mapsto S'_a{}^b \psi_b\,, ~~~~~ \psi_{\bar a} \mapsto (S'^{-1})^T{}_{\bar a}{}^{\bar b}\psi_{\bar b}\,.
\end{equation}

We must now determine exactly what choices of $S'$ correspond to our asymmetric orbifolds. It is a well known fact in algebra that any finite order automorphism of a free abelian group $\mathbb{Z}^p$ can be represented by an invertible $p\times p$ matrix with integral entries, i.e. a matrix in $GL(p,\mathbb{Z})$. In our case, the action of the orbifold symmetry on the Narain lattice $\Gamma_{4,4}$ is restricted to a sublattice of rank at most 4, which is indeed an automorphism of a free abelian group $\mathbb{Z}^4$. For $n=5,8,10,12$, the $\mathbb{Z}_n$ operation can always be represented by the companion matrix
\begin{align}
S' &= \begin{pmatrix}
0 & 0 & 0 & -a_0\\
1 & 0 & 0 & -a_1\\
0 & 1 & 0 & -a_2\\
0 & 0 & 1 & -a_3
\end{pmatrix},
\end{align}
where $a_i$ are the coefficients of the cyclotomic polynomial
\begin{align}
\Phi_n(x):=\prod_{k|n}(x-e^{2\pi i k/n})&= a_0 + a_1 x + a_2 x^2+a_3 x^3 + x^4.
\end{align}
The matrices for $n=2,3,4,6$ are block matrices where the diagonal entries are $2\times 2$ companion matrices, see Appendix \ref{app:theta}. 

At this point we should note that the group $Spin(4,4;\Z)$ contains in general various inequivalent elements with the same order $n$. We are however restricting such elements to be of the block diagonal form in eq. \eqref{S}, which are unique up to conjugation in $Spin(4,4;\mathbb{Z})$. This restriction guarantees their correspondence with the T-duality action on $\psi$ occurring in our asymmetric orbifolds.\footnote{An instructive example is as follows. Consider the case $n = 2$. The matrix $S$ realizes a spinor representation of $Spin(4,4;\mathbb{Z})$, but in a suitable basis it will realize a cospinor representation acting on RR 1-forms. The form of $S$ guarantees that its action projects out all of the eight RR 1-forms, which occurs only for the spinor representation of the order two element $M \in Spin(4,4;\mathbb{Z})$ defining the asymmetric orbifold. The order 2 element acting on $T^4$ to yield a K3 surface, for example, preserves all RR 1-forms. Yet another order 2 element preserves half of the RR 1-forms. }

We now enhance the action of $S$ on $\psi$ by an axionic shift. For our purposes it will be enough to focus on the fields $C_i$, and for simplicity we switch from the $a,b,c$ indices to $i,j,k$. The axionic shift is given by
\begin{equation}\label{0formsu}
    T(u): ~~~ C_i \mapsto C_i + u_i\,, ~~~~~ u_i \in \mathbb{Z}\,,
\end{equation}
and so the extended orbifold symmetry acts as
\begin{equation}
    gT(u): ~~~ C_i \mapsto S'_i{}^j(C_j + u_j)\,.
\end{equation}
The question is if there exists a choice of $u_i$ such that the fixed values of $C_i$ are fractional. The only other possibility is that they are null modulo integral shifts, in which case our orbifold is equivalent to the original one. In other words we look for $u_i$ such that the equation
\begin{equation}\label{fixed}
    C_i = S'_i{}^j(C_j + u_j)
\end{equation}
has a solution with $C_i \notin \Z$ for some or all $i$. As advertised, we show in Table \ref{tab:theta-solutions} that such $u_i$ exist when the order of the orbifold symmetry is $n = 2,3,4,5,8$, but not when $n = 6,10,12$. In the following Section we show that the fractional $C_i$ are identified with discrete theta angles correlated with an incompleteness of the BPS string spectrum. 

\subsubsection{Effect on the fundamental string charge lattice}

To see how the discrete theta angle affects the spectrum of strings we must determine how $gT(s)$ acts on the 2-form fields of the effective 6D theory. These fields come from the reduction of the 10D 3-form and its dual 5-form, and, similarly to the RR axions, they arrange into a T-duality spinor
\begin{equation}
    \widetilde{\psi}_{\mu\nu} = (C_{1\mu\nu},C_{2\mu\nu},C_{3\mu\nu},C_{4\mu\nu},C_{234\mu\nu},C_{134\mu\nu},C_{124\mu\nu},C_{123\mu\nu})\,,
\end{equation}
where $\mu,\nu = 0,...,5$ are the 6D Lorentz indices. Under the orbifold T-duality action, the first four components transform as
\begin{equation}\label{2formst}
    C_{i\mu\nu} \mapsto S'_i{}^{j}C_{j\mu\nu}\,,
\end{equation}
while under the axionic shift they transform as
\begin{equation}
    T(u): ~~~ C_{i\mu\nu} \mapsto C_{i\mu\nu} + u_i B_{\mu\nu}\,,
\end{equation}
where $B_{\mu\nu}$ is the Kalb-Ramond B-field. The combined action $gT(u)$ then reads 
\begin{equation}\label{2formsu}
    gT(u): ~~~ C_{i\mu\nu} \mapsto S'_i{}^j(C_{j\mu\nu} + u_jB_{\mu\nu})\,.
\end{equation}
The field $B_{\mu\nu}$ itself is left invariant under $gT(u)$. The fields $C_{ijk\mu\nu}$ are invariant under $T(u)$, hence under $gT(u)$ they are acted on by $(S'^{-1}){}^T$ and projected out as usual. 

In the absence of a discrete theta angle, e.g. when $u_j = 0$, eq. \eqref{2formsu} reduces to \eqref{2formst} and so the 2-forms $C_{i\mu\nu}$ are projected out in the same way as the RR axions $C_i$, cf. eq. \eqref{0formsu}. More generally there exist four linear combinations
\begin{equation}\label{lincom}
    \widetilde{C}^{(2)}_i = a_i{}^jC^{(2)}_j + b_i{}^ju_jB^{(2)}
\end{equation}
such that eq. \eqref{2formsu} is equivalent to
\begin{equation}\label{tor}
    \widetilde{C}^{(2)}_i = S'_i{}^j \widetilde{C}^{(2)}_j\,,
\end{equation}
where the superscripts signify the degree of the $p$-forms. These combinations are the $n$-torsional 2-forms generalizing that of the AOB example in Section \ref{ss:AOB}.

Without loss of generality we may set $a_i{}^j = \delta_i{}^j$ in \eqref{lincom}, in which case $b_i{}^j$ is constrained by eq. \eqref{tor} to 
\begin{equation}\label{bcoeff}
    b_i{}^j = (1-S'^{-1})^{-1}{}_i{}^j = -[(1 - S')^{-1} S']_i{}^j\,.
\end{equation}
This matrix is always well defined since $S'$ never has eigenvalues $\lambda = 1$. On the other hand, equation \eqref{fixed} can be written as
\begin{equation}
    C_i^{(0)} = [(1 - S')^{-1} S']_i{}^j u_j\,,
\end{equation}
and we obtain a general formula expressing the torsional 2-forms in terms of the discrete theta angles:
\begin{equation}
    \widetilde{C}^{(2)}_i = C_i^{(2)} - C_i^{(0)}B^{(2)}\,.
\end{equation}

Consider now the strings which are charged under the 2-form fields $B^{(2)}$ and $C^{(2)}_i$ in the parent theory. We rewrite their coupling in terms of the torsional 2-forms as
\begin{equation}
    \int qB^{(2)} + p^iC_i^{(2)} = \int (q + C^{(0)}_ip^i)B^{(2)} + p^i \widetilde{C}_i^{(2)}\,.
\end{equation} 
Setting $q = 0$, we see from this equation that the discrete theta angles $C_i^{(0)}$ are correlated with the presence of non-BPS strings with fractional B-field charge $C^{(0)}_ip^i$, extending the string charge lattice with sites with no BPS representative. Moreover, this extension is precisely given by the order of the discrete theta angle. 

Let us now make some comments on the spectrum of solitonic strings. It was conjectured recently in \cite{Etheredge:2025rkn} that, in the context of the lattice weak gravity conjecture (LWGC), theories where only a proper sublattice of electric charges has superextremal representatives, there are magnetically charged states that become confined which would otherwise violate the Dirac quantization condition. Such a violation of the full LWGC is realized in the string compactifications with discrete theta angles of \cite{Montero:2022vva}, and the confinement phenomenon was in fact shown to hold in \cite{Etheredge:2025rkn} for them. In the theories we are presently considering, the prediction is then that the solitonic strings which are fractionally paired with the extra non-BPS states extending the string charge lattice must become confined as the discrete theta angle is turned on. We leave an analysis of how this occurs for future work. 

\noindent 

\subsubsection{Effect on particle charge lattice}
The five new theories we have constructed in this article were predicted to exist in \cite{Fraiman:2022aik} taking as their defining datum the lattice of electric charges for particles. It was suggested in that reference that two of the string islands were characterized by having discrete theta angles turned on from the observation that their corresponding lattices were paired in a specific way with those of two other theories, in a way that already occurred for the three known theories of this type \cite{Montero:2022vva,ParraDeFreitas:2022wnz}, see Table \ref{tab:6Dlatpairs}. We will refer to the charge lattices of theories with discrete theta angle turned off and on respectively as $\Gamma_c$ and $\Gamma_c'$.

\begin{table}
    \centering
    \begin{tabular}{|c|c|c|}
    \hline
		$n$	& $\Gamma_c$&$\Gamma_c'$\\ \hline\hline
2	&		$\Gamma_{1,1}\oplus \Gamma_{3,3}(2)$ & $\Gamma_{4,4}(2)$ \\ \hline
3	&		$A_2(-1)\oplus \Gamma_{1,1}\oplus \Gamma_{1,1}(3)$ & $A_2(-1)\oplus\Gamma_{2,2}(3)$ \\ \hline
4	&		$2\,A_1(-1)\oplus \Gamma_{1,1}\oplus \Gamma_{1,1}(4)$ & $2\,A_1(-1)\oplus \Gamma_{1,1}(2)\oplus \Gamma_{1,1}(4)$ \\ \hline
5	&		$A_4(-1)$ & $A_4^*(-5)$ \\ \hline
8	&		$A_1(-1)\oplus A_3(-1)$ & $3\,A_1(-1)\oplus A_1(-2)$ \\ \hline
	\end{tabular}
	\caption{Pairs of charge lattices for the asymmetric orbifolds admitting discrete theta angle $\theta$, suggested by the classification scheme of \protect\cite{Fraiman:2022aik}. $\Gamma_c$ and $\Gamma_c'$ correspond respectively to having $\theta = 0$ and $\theta \neq 0$.}
    \label{tab:6Dlatpairs}
\end{table}

In general, turning on a discrete theta angle introduces extra non-BPS particles whose electric charges may not be in $\Gamma_c$, in which case $\Gamma_c'$ is a proper overlattice of $\Gamma_c$. We have already seen how this occurs for the $\Z_2$ asymmetric orbifold in Section \ref{ss:AOB}, where $\Gamma_c = \Gamma_{1,1}\oplus \Gamma_{3,3}(2)$ and $\Gamma_c' = \Gamma_{4,4}(2)$ in 6D with a suitable normalization. In this Section we perform the general computation of both $\Gamma_c$ and $\Gamma_c'$ for 6D asymmetric orbifolds, verifying that the results agree exactly with the output of the classification scheme of \cite{Fraiman:2022aik}. 

For technical reasons, it is easier to compute both $\Gamma_c$ and $\Gamma_c'$ in a somewhat indirect manner. Namely we will consider a construction starting from the type IIA string on $T^4\times S^1$, with $g = (M,v)$ such that $M$ acts on the $T^4$ internal CFT as before, but $v$ is taken as an order $N$ shift along the extra $S^1$. In other words, $v$ lies in the $\Gamma_{1,1}$ part of the $T^4\times S^1$ Narain lattice $\Gamma_{4,4}\oplus \Gamma_{1,1}$ instead of the $\Gamma_{4,4}$ part. The resulting theory is then T-dual to the $S^1$ compactification of the 6D theories we have already constructed --- indeed, we already used this fact in Section \ref{ss:quasicrystals} to determine the duality relationship between $S^1$ compactificaitons of type II string islands and the quasicrystallographic compactifications of \cite{Baykara:2024vss}.\\

\noindent \textbf{Computation of $\Gamma_c$:}

\noindent The 5D versions of the eight 6D asymmetric orbifolds form part of a larger family of 5D theories classified and examined in \cite{Persson:2015jka} (all without discrete theta angle). This reference includes a computation of the respective charge lattices $\Gamma_c$. We will first review briefly how to carry out this computation, and then determine how it is affected by turning on the discrete theta angle. 

We compute first the charge lattice of states in the untwisted sector of the orbifold. The T-duality symmetry $M$ acts on the Narain lattice $\Gamma_{4,4}$ of the $T^4$ part of the internal CFT, fixing the \textit{invariant lattice} 
\begin{equation}
    I\subset \Gamma_{4,4}\,.
\end{equation}
The action of $M$ on the NSNS 1-forms preserve the $U(1)$ fields whose charge is measured by the momentum operators forming the lattice $I$. There are also states charged under the surviving as well as the projected out $U(1)$'s. It is a general fact \cite{Narain:1986qm} that the projection of such states on $I$ extend this lattice to its dual $I^*$. Therefore, before enforcing the orbifold projection, the lattice of electric charges with respect to the invariant $U(1)$'s in the untwisted sector reads
\begin{equation}\label{untwlat}
     \Gamma_{c,\text{untwisted}} = I^*\oplus \Gamma_{1,1}\,,
\end{equation}
where $\Gamma_{1,1}$ corresponds to the extra $U(1)$'s coming from the $S^1$ compactification. The orbifold projection acts on both the oscillators (through $M$) and the momentum modes along $S^1$ (through $v$) by multiplication by $n$-th roots of unity. Therefore for every site in $I^*\oplus \Gamma_{1,1}$ there always exists a set of invariant states, and eq. \ref{untwlat} gives exactly the charge lattice of untwisted states.

In the twisted sectors, the translation along $S^1$ introduces a fractional shift on the winding number $w$, so that $w \in \Z + p/n$, $q = 1,2,...,n-1$ in the $g^p$-twisted sector. These states too have charges on the lattice $I^*$, hence each of the twisted sectors has a charge lattice equal to a translate of $\Gamma_{c,\text{untwisted}}$. Putting together the contributions from all the sectors in the orbifold, we obtain the full charge lattice
\begin{equation}\label{fulllat}
     \Gamma_c = I^*\oplus \Gamma_{1,1}(1/n)\,.
\end{equation}
The invariant lattices for each of the eight asymmetric orbifolds are obtained from the construction in Section \ref{ss:explicit}, and read
\begin{equation}
\begin{split}
    I &= \Gamma_{4,4}\,, ~~~~~~~~ \Gamma_{2,2}\oplus A_2(-1)\,, ~~~~~~~~ \Gamma_{2,2}\oplus 2\,A_1(-1)\,, ~~~ \Gamma_{2,2}\oplus A_2(-1)\,, \\
    &~~~~ A_4(-1)\,, ~~~ A_1(-1)\oplus A_3(-1)\,, ~~~ A_4(-1)\,, ~~~~~~~~~~~~~~ A_2(-1)\oplus 2A_1(-1)\,,
\end{split}
\end{equation}
respectively for $n = 2,3,4,5,6,8,10,12$.\\

\noindent \textbf{Computation of $\Gamma_c'$:}

\noindent In Section \ref{ss:AOB} we saw how the extra non-BPS strings wrap on cycles to produce non-BPS particle states extending the lattice $\Gamma_c$ to $\Gamma_c'$. The same conclusion can be reached by studying directly how the spectrum of particles changes due to the action of the shift $T(u)$ on RR 1-form fields, just as we did for the string spectrum. 

There are eight RR 1-forms coming from the 10D 1-form as well as reductions of the 3-form and 5-form on $T^4$ cycles, and these form a cospinor representation of the T-duality group $Spin(4,4;\Z)$,  
\begin{equation}
  \psi_{c,\mu} = (C_\mu, C_{12\mu}, C_{13\mu}, C_{14\mu};C_{1234\mu},C_{34\mu},C_{24\mu},C_{23\mu})\,.
\end{equation}
For our purposes we can choose some basis in which the T-duality transformation $M$ is represented by $S$ in eq. \eqref{S}. In this basis, the spinors of 0-forms and 2-forms will transform differently, but this will not affect the present computation. The axionic shift $T(u)$ acts on the 1-form fields as
\begin{equation}\label{coshift}
  T(u): ~~~~~ C_\mu \mapsto C_{\mu}\,, ~~~ C_{1234\mu} \mapsto C_{1234\mu}\,, ~~~ C_{ij\mu} \mapsto C_{ij\mu} + u_i B_{j\mu} + u_j B_{i\mu}\,.
\end{equation}
We define $\chi_\mu$ such that $T(u): \psi_{c,\mu} \mapsto \psi_{c,\mu}+\chi_\mu$, and so the full orbifold operation $gT(u)$ reads
\begin{equation}
  gT(u): ~~~~~ \psi_{c,\mu}\mapsto S(\psi_{c,\mu} + \chi_\mu)\,.
\end{equation} 
Similarly to the 2-form fields in the previous Section, cf. eqs. \eqref{lincom} to \eqref{bcoeff}, the torsional 1-forms can be written as 
\begin{equation}
    \widetilde{\psi}_{c,\mu} = \psi_{c\mu} + \delta_\mu\,, ~~~~~\delta_\mu \equiv - (1-S)^{-1}S\chi_\mu\,.
\end{equation}
Note that in this case all of the eight 1-forms, and not only four, are affected in a nontrivial way by the shift $T(u)$ --- even if $C_\mu$ and $C_{1234\mu}$ are invariant under $T(u)$, the combination $gT(u)$ mixes all of the eight degrees of freedom of $\psi_{c,\mu}$. 

The shift $\chi_\mu$ is a function of the 1-forms $B_{i\mu}$, and so it is correlated with extra non-BPS particles with fractional charges under these fields. To be precise, consider the coupling of particles to the NSNS 1-forms $B_{i\mu}$ and RR 1-forms $\psi_{c,\mu}$, written in terms of the torsional RR 1-forms $ \widetilde{\psi}_{c,\mu}$,
\begin{equation}
    \int w^iB_{i\mu} + m^\alpha\psi_{c,\mu\alpha} = \int w^iB_{i\mu} + m^\alpha(\widetilde{\psi}_{c,\mu\alpha}-\delta_{\mu\alpha})\,.
\end{equation}
where $\alpha$ is the cospinor index, and $m^\alpha$ encodes the charges of particle states coming from 0-branes, 2-branes and 4-branes suitably wrapped on $T^4$ cycles. In general the components of $\delta_\mu$ are linear combinations of the $B_{i\mu}$ with fractional coefficients:
\begin{equation}
    \delta_\mu = r^i B_{i\mu}\,,
\end{equation}
where $mr^i \in \mathbb{Z}$ with $m$ the order of the discrete theta angle. It can be verified for each of the five orbifolds $\Z_{2,3,4,5,8}$ that the overall effect is to extend, for each of the four fields $B_{i\mu}$, the lattice of charges such that the minimal charge is now $1/m$ instead of $1$. 

This effect is computationally equivalent to allowing fractional windings along each of the four compact directions, hence prior to the orbifold projection, the Narain lattice $\Gamma_{4,4}$ is extended to $\Gamma_{4,4}(1/m)$. After projecting out states, then, the overall change in the charge lattice $\Gamma_c$ due to the discrete theta angle is a scaling $I^* \mapsto I^*(1/m)$, and so
\begin{equation}
    \Gamma_c' = I^*(1/m) \oplus \Gamma_{1,1}(1/n)\,,
\end{equation}
cf. eq. \eqref{fulllat}. We should note that there are also extra particle states arising from the extra non-BPS strings wrapped around $S^1$. However, it is always the case that $n/m \in \mathbb{Z}$, hence the charges of these extra states always have representatives in $\Gamma_{1,1}(1/n)$ coming from BPS strings with with fractional winding along the same $S^1$. Indeed we already saw this happening for the $\Z_2$ orbifold in Section \eqref{ss:AOB}. 

We record the computation of the five lattices $\Gamma_c'$ in Table \ref{tab:5Dlats}, where we have scaled them by a factor $mn$ to $I^*(n)\oplus \Gamma_{1,1}(m)$ in order to make them even and to compare them with the predictions made in \cite{Fraiman:2022aik}. Quite remarkably, we see an exact match, once we account for the fact that we are comparing between theories constructed with shifts along different parts of the Narain lattice. These shifts are ultimately equivalent, as verified by the fact that the charge lattices in both columns are globally isomorphic, i.e. they describe the same theory at different points in moduli space. 

\begin{table}
    \center
    \begin{tabular}{|c|c|c|}
    \hline
		$n$	& Predicted $\Gamma_{c,\text{5D}}'$ \cite{Fraiman:2022aik}& Computed $\Gamma_{c,\text{5D}}'$\\ \hline\hline
2	&		$\textcolor{Red}{\Gamma_{4,4}(2)}\oplus \Gamma_{1,1}(2)$ & $\Gamma_{4,4}(2)\oplus \textcolor{Red}{\Gamma_{1,1}(2)}$ \\ \hline
3	&		$\textcolor{Red}{A_2(-1)\oplus \Gamma_{2,2}(3)}\oplus \Gamma_{1,1}(3)$ & $A_2(-1)\oplus \Gamma_{2,2}(3)\oplus \textcolor{Red}{\Gamma_{1,1}(3)}$ \\ \hline
4	&		$\textcolor{Red}{2\,A_1(-1)\oplus \Gamma_{1,1}(2)\oplus \Gamma_{1,1}(4)}\oplus \Gamma_{1,1}(4)$ & $2\,A_1(-1)\oplus \Gamma_{2,2}(4)\oplus \textcolor{Red}{\Gamma_{1,1}(2)}$ \\ \hline
5	&		$\textcolor{Red}{A_4^*(-5)}\oplus \Gamma_{1,1}(5)$ & $A_4^*(-5)\oplus\textcolor{Red}{\Gamma_{1,1}(5)}$ \\ \hline
8	&		$\textcolor{Red}{3A_1(-1)\oplus A_1(-2)}\oplus \Gamma_{1,1}(8)$ & $A_3^*(-8)\oplus A_1(-2)\oplus \textcolor{Red}{\Gamma_{1,1}(2)}$ \\ \hline
		\end{tabular}
	\caption{Charge lattices $\Gamma_{c\text{5D}}'$ of the 5D versions of our asymmetric orbifolds with discrete theta angle $\theta \neq 0$. The predicted lattices are given by those of the 6D theories in Table \ref{tab:6Dlatpairs} labeled $\Gamma_c'$ with an extra factor $\Gamma_{1,1}(n)$ associated to a trivial $S^1$ compactification down to 5D. The red coloring indicates the sublattice to which the shift vector is restricted. The computed lattices result from constructing the asymmetric orbifolds of the type IIA string on $T^4\times S^1$ with the shift given by an order $n$ translation along the extra circle, as denoted by the red coloring. In all five cases, the computed lattices agree with the predicted ones. The case $n=8$ requires the use of the global lattice isomorphism $3A_1(-1)\oplus A_1(-2) \Gamma_{1,1}(8)\simeq A_3^*(-8)\oplus A_1(-2) \oplus \Gamma_{1,1}(2)$, which can be verified using \texttt{Magma} or \texttt{SageMath}. } \label{tab:5Dlats}
\end{table}

\section{Discussion of results}\label{s:conclusions}
In this paper we have constructed explicitly the type II asymmetric orbifolds (as well as their versions with discrete theta angle) predicted to exist in \cite{Fraiman:2022aik} from a formal rank reduction map. Together with the recent results of \cite{Aldazabal:2025zht} for heterotic strings, this completes the predicted list of cyclic asymmetric orbifolds. These latter results however point to a subtle caveat. Namely, it is possible that there are two heterotic islands and not one. These results are not contradictory to our own, as we now explain.

It was argued in \cite{Fraiman:2022aik} that the predicted cyclic orbifolds are in one-to-one correspondence with the classification of genera of 2D chiral bosonic CFTs with $c = 24$ \cite{Schellekens:1992db,Hohn:2017dsm,Hohn2022} in the heterotic case and currents of the 2D chiral sCFT comprising 24 fermions in the type II case \cite{Harrison:2020wxl}. We have given in this paper a constructive proof that this relationship does hold, although not necessarily in a one-to-one manner. Our proof admits the possibility that for any 2D CFT there may correspond more than one 6D theory depending on the genus of the invariant lattice for construction of the latter. This is exactly the case of the heterotic island of \cite{Aldazabal:2025zht}, where the genus of invariant lattices contains two representatives, and the authors constructed a well defined asymmetric orbifold for each of them. Our results imply that, if these theories are genuinely inequivalent, they must become T-dual under circle compactification on $S^1$ to 5D. 

On the other hand, our proof excludes the necessity of accounting for non-cyclic orbifolds --- they, too, must correspond to chiral 2D CFTs. In the case that a non-cyclic orbifold can be constructed consistently, it must be T-dual to a cyclic one. An example is the $\Z_2\times \Z_2$ asymmetric orbifold of the heterotic string on $T^4$ \cite{deBoer:2001wca}; it can be obtained by first performing a $\Z_2$ orbifolding which reduces the rank by 8, and then another $\Z_2$ reducing the rank further by 4. We then expect this non-cyclic orbifold to be equivalent to the cyclic one with rank reduced by 12. This latter orbifold was constructed in \cite{Aldazabal:2025zht} and is in correspondence with the set of 2D chiral bosonic CFTs comprising the so-called genus D of \cite{Hohn:2017dsm}. One can see for example from Figure 4 of \cite{Hohn:2023auw} that these latter CFTs can be obtained either as $\Z_2$ orbifolds of theories in genus A (corresponding to the unorbifolded heterotic string on $T^4$) or as $\Z_2$ orbifolds of theories in genus B (corresponding to the theory with rank reduced by 8), exactly in accordance with our prediction. 

Our results then rule out the 23 extra theories suggested to exist in \cite{Fraiman:2022aik} from an iterative application of the rank reduction map. A motivation for proposing these theories was the existence of the $\Z_2 \times \Z_2$ asymmetric orbifold above, but as we have just argued, this orbifold fits into the classification of cyclic ones! This clue turned out to be, in fact, a red herring. To clarify why the rank reduction map cannot be applied iteratively in a naive manner, we require a physical explanation for its existence. This explanation will appear in a future work \cite{RR}, where it will be shown exactly how this iterative application is physically obstructed. 

It is also remarkable that there exist string islands with nontrivial discrete theta angle. Because these theories have a BPS-incomplete lattice of both strings and particles, they violate the lattice weak gravity conjecture (LWGC) of \cite{Heidenreich:2016aqi} for both spectra, and satisfy instead a \textit{sublattice} constraint (sLWGC). One may ask what are the general properties of the theories which violate the LWGC. It was recently proposed in \cite{Etheredge:2025rkn} that one such property is the confinement of states in the dual magnetic lattice which violate Dirac quantization with respect to the sites violating the LWGC, and in particular it was shown that this phenomenon does occur in the string vacua with discrete theta angle of \cite{Montero:2022vva}. In aiming to determine what are the \textit{universal} consequences of violating the LWGC, it is desirable to have examples with the fewest possible moduli; this narrows down the range of effects which are moduli-dependent and therefore allow for the most stringent tests. Our string islands are so far the best such examples, and their analysis in this context should be very instructive. 

Let us also note that the $\Z_5$ island violates the LWGC for strings in the strongest way as measured by the \textit{coarseness} of the sublattice of charges for which all sites have superextremal representatives. This coarseness is equal to the order of the discrete theta angle, which is 5 for the $\Z_5$ island. At least for theories with 16 supercharges, the previously known examples had coarseness 2 and 3. 

Finally, we have shown that the type II string islands, compactified on an extra $S^1$, are related to the 5D quasicrystalline compactifications of \cite{Baykara:2024vss} as Sen-Vafa dual pairs \cite{Sen:1995ff}, clarifying the role of the latter in describing the string landscape. Since this map is a strong-weak relation, it is expected that an interesting decompactification of the quasicrystal will exist that corresponds to the S-dual of the 6d string island which has been left for future work. What we have not examined are the effects of turning on the discrete theta angle in the dual quasicrystal. It would be very interesting to study these effects, specially in light of the confinement phenomenon of \cite{Etheredge:2025rkn}.

\section*{Acknowledgements}

We thank Muldrow Etheredge, Anamaria Font, Bernardo Fraiman, Miguel Montero and Cumrun Vafa for useful discussions and correspondence. HCT thanks Harvard University and the Swampland Initiative for its hospitality. The work of HP and ZKB is supported by a grant from the Simons Foundation (602883,CV), the DellaPietra Foundation and by the NSF grant PHY2013858. 

\appendix

\section{Classification of 6D $\mathcal{N} = (1,1)$ asymmetric orbifolds}\label{app:class}
In this Appendix we wish to prove that asymmetric orbifolds of the type II string on $T^4$ of the type $(0,2)$, i.e. preserving two right-moving gravitini, are always in correspondence with orbifolds of the chiral superconformal field theory based on the $E_8$ lattice preserving its $\mathcal{N} = 1$ structure. The latter are realized as systems of 24 free fermions supporting a current algebra of dimension 24, of which there are eight choices, and so the full set of type II asymmetric orbifolds of the stated form comprises eight different theories in one-to-one correspondence with these current algebras. These are precisely the eight theories discussed in the text. 

To prove this statement, we will rely on the underlying momentum lattice structure of the worldsheet CFTs and exploit various lattice theoretical results of Nikulin \cite{NikulinVV1980ISBF}. Given the generality of our result and its strong implications, we will try to keep an appropriate level of mathematical rigor, making this appendix rather technical but hopefully still readable to non-experts. We will start in Section \ref{app:prel} with some preliminary lattice theory results of a general nature and then go through the steps of building a proof of the advertised classification result in subsection \ref{app:class}. The reader may find it useful to complement our discussion with that of \cite{Volpato:2014zla}, where similar techniques are explained in detail. For standard definitions, notation, and lattice basics (e.g. primitive embeddings) we refer the reader to \cite{NikulinVV1980ISBF}.

\subsection{Preliminaries}\label{app:prel}
\subsubsection{Three results of Nikulin}
We will need the following results of Nikulin:\\

\noindent \textbf{Theorem 1:} (\textit{Theorem 1.12.4 of \cite{NikulinVV1980ISBF}}). Let there be given two pairs of nonnegative integers, \( (t_{(+)}, t_{(-)}) \) and \( (l_{(+)}, l_{(-)}) \). The following properties are equivalent:

\begin{enumerate}
    \item[a)] Every even lattice of signature \( (t_{(+)}, t_{(-)}) \) admits a primitive embedding into some even unimodular lattice of signature \( (l_{(+)}, l_{(-)}) \).
    \item[b)] \( l_{(+)} - l_{(-)} \equiv 0 \pmod{8}, \, t_{(+)} \leq l_{(+)}, \, t_{(-)} \leq l_{(-)} \) \\
    and \( t_{(+)} + t_{(-)} \leq \frac{1}{2}(l_{(+)} + l_{(-)}) \).
\end{enumerate}

\noindent \textbf{Theorem 2:} (\textit{Proposition 1.6.1 of \cite{NikulinVV1980ISBF}}). A primitive embedding of an even lattice $S$ into an even unimodular lattice, in which the orthogonal complement of $S$ is isomorphic to $K$, is determined by an isomorphism $\gamma: A_S \overset{\sim}{\rightarrow} A_K$ for which $q_K \circ \gamma = -q_S$.\\

\noindent \textbf{Theorem 3:} (\noindent\textit{Corollary 1.13.4 of \cite{NikulinVV1980ISBF}}). Let \( T \) be an even lattice with invariants \( (t_{(+)}, t_{(-)}, q) \). Then \( \Gamma_{1,1} \oplus T \) is the unique even lattice with invariants \( (1 + t_{(+)}, 1 + t_{(-)}, q) \).\\

\subsubsection{Two required lemmas}
Now we use the above results to prove two lemmas to be used in the main proof of our classification theorem for asymmetric orbifolds.\\

\noindent \textbf{Lemma 1:} Let $I_{n,4}$ be an even lattice with signature $(n,4)$. There always exists a lattice $I_{4+n}$ with signature $(4+n,0)$ such that there is a global isomorphism
\begin{equation}\label{proof1}
    I_{n,4} \oplus \Gamma_{4,4} \simeq I_{4+n} \oplus E_8(-1)\,.
\end{equation}

\noindent \underline{\textit{Proof:}} First note that by Theorem 1 the lattice $I_{n,4}$ admits a primitive embedding into $\Gamma_{n,n+8} \simeq\Gamma_{n,n}\oplus E_8$. By Theorem 2, the orthogonal complement of this primitive embedding is a lattice with signature $(0,4+n)$ with discriminant form $q' = -q$ where $q$ is the discriminant form of $I_{n,4}$. Define $I_{4+n}$ as the associated lattice with bilinear form multiplied by $-1$; its discriminant form is $q$. Since $E_8(-1)$ and $\Gamma_{4,4}$ have trivial discriminant form, both the LHS and RHS of \eqref{proof1} have discriminant form $q$ hence they are locally isomorphic. Since the LHS has sublattice $\Gamma_{4,4}$, theorem 3 implies that the isomorphism is global.  \hfill $\blacksquare$\\

\noindent \textbf{Lemma 2:} Let $I_{n,4}$ and $I_{4+n}$ be as above. Let $v$ be a vector in the rational span of $I_{n,4}$, $I_{n,4}\otimes_{\mathbb{Z}} \mathbb{Q}$, such that $nv \in I_{n,4}$ for some $n \in \Z$. There exists a vector $u \in I_{n,4}\oplus \Gamma_{4,4}$ such that $v' = v+u$ is in $I_{4+n}\otimes_{\mathbb{Z}} \mathbb{Q}$ in the RHS of \eqref{proof1}.\\

\noindent \underline{\textit{Proof:}} The vector $v$ by construction is such that $v \notin I_{n,4}^*$, hence it defines a nontrivial even sublattice
\begin{equation}
    I_{n,4}^{(v)} \equiv \{u\in I_{n,4} | u\cdot v \in \Z\}\,.
\end{equation}
By Lemma 1 there exists $I_{4+n}^{(v')}$ such that 
\begin{equation}
    I_{n,4}^{(v)}\oplus \Gamma_{4,4} \simeq I_{4+n}^{(v')}\oplus E_8(-1)\,,
\end{equation}
and 
\begin{equation}
    I_{4+n}^{(v')} \equiv \{u \in I_{4+n}| u \cdot v' \in \Z \}
\end{equation}
for some vector $v' \in I_{4+n}\otimes_{\mathbb{Z}} \mathbb{Q}$. It follows that $v'$ and $v$ define globally isomorphic sublattices of $I_{n,4}\oplus \Gamma_{4,4}$, hence they are in the same conjugacy class of the discriminant group of $I_{n,4}^{(v)}\oplus \Gamma_{4,4}$. This implies that $v' = v + u$ for $u \in I_{n,4}^{(v)}(-1)\oplus \Gamma_{4,4} \subset I_{n,4}\oplus \Gamma_{4,4}$. \hfill $\blacksquare$\\

\subsection{Proof of classification theorem}\label{app:proof}
We now move on to prove the proposition stated in Section \ref{ss:classification}:\\

\noindent\fbox{
\parbox{\textwidth}{%
\vspace{0.1in}
\textbf{Proposition:} Any asymmetric orbifold of the type II string on $T^4$ by a symmetry group $\mathcal{G}$ preserving the $\mathcal{N}=(1,1)$ structure and acting trivially on right (or left) moving gravitini is in correspondence with an orbifold of the $E_8$ sCFT by some symmetry group $\mathcal{G}'$ which acts trivially on a sublattice of rank at least 4 and preserves the $\mathcal{N} = 1$ structure. This correspondence is realized as an equivalence between the $T^4$ compactified type II orbifold and the tensor product of the $E_8$ sCFT orbifold with an antiholomorphic $E_8$ sCFT.
\vspace{0.1in}
}
}\\

\noindent We will start by proving this result at the level of cyclic orbifolds, which establishes the uniqueness of the 6D cyclic asymmetric orbifolds presented in this paper. We then show how this result extends to the general symmetry groups $\mathcal{G}$ and $\mathcal{G'}$ including non-cyclic cases.

\subsubsection{Cyclic case}
Let $g = (M,v)$ be a global symmetry of the type IIA string compactified on $T^4$ defining an asymmetric orbifold as explained in Section \ref{ss:asymorb}. The symmetry $M$ acts on the Narain lattice $\Gamma_{4,4}$ defining the \textit{invariant sublattice} $I_{n,4}$ with $n \geq 0$. Compactifying the 6D theory on $T^4$ as in Section \eqref{ss:classification}, the Narain lattice extends to $\Gamma_{4,4}\oplus \Gamma_{4,4} \simeq \Gamma_{8,8}$, and the invariant lattice reads
\begin{equation}
    I_{n+4,8} \simeq I_{n,4}\oplus \Gamma_{4,4}\,.
\end{equation}
By \textbf{Lemma 1} there exists a lattice boost under which the invariant lattice transforms as
\begin{equation}\label{invtransf}
    I_{n,4}\oplus \Gamma_{4,4} \mapsto I_{4+n}\oplus E_8(-1)\,,
\end{equation}
with $I_{4+n}$ a positive definite lattice. The Narain lattice necessarily transforms to
\begin{equation}\label{typeIIpol}
    \Gamma_{8,8} \mapsto E_8 \oplus E_8(-1)\,,
\end{equation}
bringing the type IIA worldsheet in 2D to its holomorphic factorization point. The worldsheet here consists of two copies of the (anti)holomorphic $\mathcal{N} = 1$ $E_8$ sCFT, in such a way that the action of $M$ defines an invariant lattice $I_{4+n}\subset E_8$. It is also clear that, having started from a symmetry acting on $E_8$ in two dimensions, the procedure can be reversed to make it act on $\Gamma_{4,4}$ only. In other words, there is a correspondence between the T-duality symmetries of the $T^4$ $\mathcal{N} = (1,1)$ worldsheet and those of the $\mathcal{N} = 1$ $E_8$ sCFT. Indeed, in \cite{Volpato:2014zla} it was shown that the discrete symmetries of the $T^4$ model were in correspondence with conjugacy classes of even Weyl transformations of the $E_8$ lattice, a result that we have just recovered. 

Consider now the full action of $g = (M,v)$ on the $T^4$ model. The shift vector $v$ is associated to a phase $\exp(2\pi i v \cdot P)$ where $P = (P_R,P_L)$ is the momentum operator. In the 1-loop partition function of the asymmetric orbifold, the untwisted sector is given by the insertion of the projector $(1+g+g^2+...+g^{N-1})/N$ in the trace over the parent theory's Hilbert space, and the term corresponding to the insertion of $g$ contains the factor
\begin{equation}\label{cycproof}
    \sum_{P \in I_{n,4}} q^{\tfrac12 P_L^2} \bar q^{\tfrac12 P_R^2} e^{2\pi i v \cdot P}\,.
\end{equation}
Since $I_{n,4}$ is fixed by $M$, the marginal deformations in the parent theory corresponding to boosts of $I_{n,4}$ survive as scalar fields in the untwisted sector. From the analysis above we know that compactifying on $T^4$ down to two dimensions there exists a marginal deformation acting as
\begin{equation}\label{cycproof2}
    \sum_{P \in I_{n,4}\oplus \Gamma_{4,4}} q^{\tfrac12 P_L^2} \bar q^{\tfrac12 P_R^2} e^{2\pi i v \cdot P} \mapsto \sum_{P \in I_{4+n}\oplus E_8(-1)} q^{\tfrac12 P_L^2} \bar q^{\tfrac12 P_R^2} e^{2\pi i v \cdot P}\,,
\end{equation}
where, with a slight abuse of language, we use the same notation $P_L,P_R,v$ for the $I_{n,4}$ vectors in \eqref{cycproof} as well as for the $I_{n,4}\oplus\Gamma_{4,4}$ vectors. By Lemma 2 there exists a vector $u \in I_{n,4}\oplus \Gamma_{4,4}$ such that $v' = v + u$ lies in $I_{4+n}$, hence the phase in the LHS of \eqref{cycproof2} is equivalent to another phase acting only on states with momenta in $I_{4+n}$, thus
\begin{equation}\label{cycproof3}
    \sum_{P \in I_{4+n}\oplus E_8(-1)} q^{\tfrac12 P_L^2} \bar q^{\tfrac12 P_R^2} e^{2\pi i v \cdot P} = \sum_{P \in I_{4+n}} q^{\tfrac12 P_L^2} e^{2\pi i v' \cdot P_L} \times \sum_{P_R \in E_8(-1)} \bar q^{\tfrac12 P_R^2}\,.  
\end{equation}
It seems then that the symmetry $g' = (M,v')$ acts in a completely holomorphic way at the factorization point of the parent theory. 

To see that $g'$ truly acts holomorphically, recall that $M$ acts trivially on the right-moving states of the worldsheet comprising the spacetime gravitini. The RHS of \eqref{cycproof3} is then accompanied by spacetime characters completing the second sum over $E_8(-1)$ to the 1-loop partition function of the antiholomorphic $E_8$ sCFT. This factorization of sums appears for all $g^p$ insertions, $p = 1,...,N-1$, and since the partition function of the $E_8$ sCFT is modular-invariant, the partition function of the whole theory is holomorphically factorized. 

\subsubsection{General case}
We have just shown that symmetries $g = (M,v)$ of the form explained in Section \ref{ss:asymorb} are equivalent to symmetries $g'$ acting holomorphically in the 2D theory obtained by $T^4$ compactification at the holomorphic factorization point. In other words, the equivalence of global symmetries of the Type IIA $T^4$ model and the $\mathcal{N} = 1$ $E_8$ sCFT uplifts from T-duality symmetries to include elements in the eight $U(1)$ currents. Cyclic orbifolds of one theory are in one to one correspondence with cyclic orbifolds of the other. 

We wish to extend this result to non-cyclic orbifolds. Consider then an orbifold group $\mathcal{G}$ with two generators $g$ and $h$ each one of the form used above. The question is if these two symmetries can be simultaneously brought to act holomorphically on the 2D factorized point. This can be seen at the level of the T-duality symmetries from the fact that both $g$ and $h$ leave invariant sublattices $I_{p,4}^g$ and $I_{q,4}^h$, respectively. The marginal deformations surviving the orbifold procedure correspond to boosts of the intersection 
\begin{equation}
    I_{n,4} = I_{p,4}^g\cap I_{q,4}^h\,.
\end{equation}
As before we may take a sublattice $S_4(-1) \subset I_{n,4}$ and perform marginal deformations to boost $I_{n,4}$ to a positive definite lattice $I_{4+n}$ ad the holomorphic factorization point of the parent theory. Since the anti-holomorphic $E_8$ sCFT is unaffected, $\mathcal{G}$ must act holomorphically at this point. This result is also in agreement with \cite{Volpato:2014zla}.

Assume now that $g$ and $h$ include non-trivial gauge shifts $v_g$ and $v_h$. We must show that there exists $u_g$ and $u_h$ in $I_{n,4}\oplus \Gamma_{4,4}$ such that $v_g' = v_g + u_g$ and $v_h' = v_h + u_h$ are in the span of $I_{4+n}$. This follows from choosing $v_g$ and $v_h$ both to lie in the span of the overall invariant lattice $I_{n,4}$. For each of the shifts we may separately use \textbf{Lemma 2} as in the cyclic case above. It follows that the action of $\mathcal{G}$ is restricted to act on the holomorphic $E_8$ sCFT at the holomorphic factorization point, thus defines a freely acting orbifold of that theory. \hfill $\blacksquare$

\subsection{Extension to heterotic strings and possible degeneracies}
The logic of the previous proof extends in a straightforward way to the setup of asymmetric orbifolds of heterotic strings on $T^4$. Instead of using as a starting point the type II Narain lattice $\Gamma_{4,4}$, we use the heterotic Narain lattice $\Gamma_{20,4}$. The symmetry $M$ acts on $\Gamma_{20,4}$ defining an invariant sublattice $I_{n,4}$ just as before and the same reasoning goes through. The main difference is that, instead of the polarization \eqref{typeIIpol} we have
\begin{equation}\label{hetpol}
    \Gamma_{20,4} \mapsto N_I \oplus E_8(-1)\,,
\end{equation}
where $N_I$ is one of the 24 even self-dual Euclidean lattices of rank 24, known as Niemeier lattices. The relationship between the Weyl group of $E_8$ and the symmetries of the type II string on $T^4$ is mirrored in this case by a relationship between the automorphisms of the Niemeier lattices and the symmetries of the heterotic string on $T^4$, which is in fact an older and more well known result (see \cite{Cheng:2016org} and references therein). The fact that $N_I$ are even self-dual, just as $E_8$, allows to carry out the same steps above to show that a consistent choice of shift in the 2D holomorphic orbifold is correlated with a consistent choice of shift in the 6D theory. 

At the points in moduli space where the Narain lattice is polarized as in \eqref{hetpol}, the worldsheet CFT is holomorphically factorized with the left-moving part corresponding to a lattice CFT based on $N_I$. Application of the allowed symmetries $M$ together with appropriate shifts $v$, in a systematic way, recovers the classification of chiral bosonic CFTs with central charge $c = 24$ \cite{Schellekens:1992db,Hohn2022}. This establishes the relationship between these latter theories and the 6D $\mathcal{N} = (1,1)$ heterotic asymmetric orbifolds. 

We must be careful, however, in noting that \eqref{invtransf} is not a one-to-one relation. Its importance is in establishing a relation between some 6D with an extra $T^4$ setup and some 2D setup with holomorphic factorization. The degeneracy in this relationship is encoded in the number of representatives of the genera of $I_{n,4}$ and $I_{4+n}$. For type II strings it turns out that all of these genera have only one representative (see Table 3 of \cite{Persson:2015jka}), hence the relationship between 6D orbifolds and 2D CFTs is one-to-one. As emphasized in \cite{Aldazabal:2025zht}, however, there is one invariant lattice $I_{n+4}$ in a heterotic orbifolds whose genus has two representatives, namely the one corresponding to the heterotic island. This means that the 6D-2D relationship is a priori two-to-one.\footnote{When referring to the 2D chiral bosonic CFTs we implicitly mean their genera. The reason that this is not too relevant is that the holomorphically factorized heterotic worldsheets live all in the same moduli space and it is appropriate to coarse grain the list of chiral CFTs into their genera. In higher dimensional setups, as we discuss, this is not necessarily the case.} What this means is that at the level of the parent theory there are two distinct regions in moduli space which are not dual to each other, nor can they be continuously connected (they are points in this case), both of which can be orbifolded consistently with the ``same" symmetry $M$.

Upon circle compactification, the two invariant lattices get extended by $\Gamma_{1,1}$ and from \textbf{Theorem 3} they become globally isomorphic. In other words, the moduli introduced by the circle compactification give rise to a connected region in moduli space, including the two points above, where the vacua have the symmetry $M$. This implies that if the two heterotic islands in 6D are genuinely inequivalent, they must become T-dual upon circle compactification. Hence the degeneracy in the relationship with the 2D chiral CFTs disappears in 5D. This is exactly the same as for the unorbifolded heterotic strings, where there is a degeneracy between the 10D theories and genus A of \cite{Hohn:2017dsm}, which disappears in 9D due to heterotic T-duality.    

\section{Computation of discrete theta angles}\label{app:theta}
We look for solutions to
\begin{align}\label{eq:theta-eq}
C_i = S_i'^j(C_j+u_j),
\end{align}
with $C_j\not\in \mathbb Z$ and $u_j\in \mathbb Z$. Here, $S'$ is the matrix implementing a $\mathbb Z_n$ symmetry on a 4d lattice. For integers with Euler totient function value $\phi(n)=4$ (i.e. $n=5,8,10,12$), $S'$ is given by the companion matrix of the $n$th cyclotomic polynomial with the twist vector $\phi$ as
\begin{align}
\phi=\left(\frac 1 n,\frac p n\right)\longleftrightarrow S'_n &= \begin{pmatrix}
0 & 0 & 0 & -a_0\\
1 & 0 & 0 & -a_1\\
0 & 1 & 0 & -a_2\\
0 & 0 & 1 & -a_3
\end{pmatrix},
\end{align}
where $p$ satisfies $1<p<n-1$ and is coprime with $n$, and $a_i$ are the coefficients of the cyclotomic polynomial
\begin{align}
\Phi_n(x):=\prod_{k|n}(x-e^{2\pi i k/n})&= a_0 + a_1 x + a_2 x^2+a_3 x^3 + x^4.
\end{align}
For integers with $\phi(n)=2$ (i.e. $n=3,4,6$), the companion matrix is a $2\times 2$ matrix
\begin{align}
S''_n &= \begin{pmatrix}
0 & -a_0\\
1 & -a_1
\end{pmatrix},
\end{align}
and for $n=2$ we use a reflection matrix in 2d, $S_2'':=-I_2$. A lattice symmetry in 4d corresponding to a twist vector $\phi=(1/n,1/m)$ with $n,m=2,3,4,6$ can then be written as
\begin{align}
\phi=\left(\frac 1 n, \frac 1 m\right) \longleftrightarrow\begin{pmatrix} 
S''_n & 0\\
0 & S''_m
\end{pmatrix}.
\end{align}

In Table \ref{tab:theta-solutions}, we construct the $S_n'$ matrices as described above and present the solutions to eq. \ref{eq:theta-eq} in terms of the shift $u_j$. We see that for $n=2,3,4,5,8$, the solution for $C_i$ involves a fraction $\frac 1 p$, which means we can choose $u_j$ so that $C_i\not\in \mathbb Z$.
\begin{table}[h!]
\centering
\begin{tabular}{|c|c|c|}
\hline
$\mathbb Z_n$ & $S'_n$ & $C_i$ \\\hline
$\mathbb Z_2$ & $\begin{pmatrix}-1 & 0 & 0 & 0\\ 0 & -1 & 0 & 0\\ 0 & 0 & -1 & 0 \\ 0 & 0 & 0 & -1\end{pmatrix}$ & $\frac 1 2 (-u_1,-u_2,-u_3,-u_4)$\\
$\mathbb Z_3$ & $\begin{pmatrix}0 & -1 & 0 & 0\\ 1 & -1 & 0 & 0\\ 0 & 0 & 0 & -1 \\ 0 & 0 & 1 & -1\end{pmatrix}$ & $\frac 1 3 (-u_1-u_2,u_1-2u_2,-u_3-u_4,u_3-2u_4)$\\
$\mathbb Z_4$ & $\begin{pmatrix}0 & -1 & 0 & 0\\ 1 & 0 & 0 & 0\\ 0 & 0 & 0 & -1 \\ 0 & 0 & 1 & 0\end{pmatrix}$ & $\frac 1 2 (-u_1-u_2,u_1-u_2,-u_3-u_4,u_3-u_4)$\\
$\mathbb Z_5$ & $\begin{pmatrix}0 & 0 & 0 & -1\\ 1 & 0 & 0 & -1\\ 0 & 1 & 0 & -1 \\ 0 & 0 & 1 & -1\end{pmatrix}$ & $\begin{array}{c}\frac 1 5 (-u_1-u_2-u_3-u_4,3 u_1-2 u_2-2 u_3-2 u_4,\\
2 u_1+2 u_2-3 u_3-3 u_4,u_1+u_2+u_3-4 u_4)\\\end{array}$\\
$\mathbb Z_6$ & $\left(
\begin{array}{cccc}
 0 & -1 & 0 & 0 \\
 1 & 1 & 0 & 0 \\
 0 & 0 & 0 & -1 \\
 0 & 0 & 1 & 1 \\
\end{array}
\right)$ & $(-u_1-u_2,u_1,-u_3-u_4,u_3)$\\
$\mathbb Z_8$ & $\left(
\begin{array}{cccc}
 0 & 0 & 0 & -1 \\
 1 & 0 & 0 & 0 \\
 0 & 1 & 0 & 0 \\
 0 & 0 & 1 & 0 \\
\end{array}
\right)$ & $\begin{array}{c}\frac 1 2 (-u_1-u_2-u_3-u_4,u_1-u_2-u_3-u_4,\\u_1+u_2-u_3-u_4,u_1+u_2+u_3-u_4)\end{array}$\\
$\mathbb Z_{10}$ & $\left(
\begin{array}{cccc}
 0 & -1 & 0 & 0 \\
 1 & 1 & 0 & 0 \\
 0 & 0 & 0 & -1 \\
 0 & 0 & 1 & 1 \\
\end{array}
\right)$ & $\left(-u_1-u_2-u_3-u_4,u_1,-u_3-u_4,u_1+u_2+u_3\right)$\\
$\mathbb Z_{12}$ & $\left(
\begin{array}{cccc}
 0 & 0 & 0 & -1 \\
 1 & 0 & 0 & 0 \\
 0 & 1 & 0 & 1 \\
 0 & 0 & 1 & 0 \\
\end{array}
\right)$ & $\left(-u_1-u_2-u_3-u_4,-u_2-u_3-u_4,u_1+u_2,u_1+u_2+u_3\right)$\\
\hline
\end{tabular}
\caption{The solutions for $C_i = S_i'^j (C_j+u_j)$ in terms of the shift $u_j$. Whenever there is a fraction $\frac 1 p$ in front of the solution, we can choose $u_j$ so that $C_i$ are fractional, implying there is a solution with discrete theta angle.}
\label{tab:theta-solutions}
\end{table}

\bibliographystyle{JHEP}
\bibliography{islands}

\end{document}